\documentclass{article}

\usepackage{arxiv}

\usepackage[utf8]{inputenc} 
\usepackage[T1]{fontenc}    
\usepackage{hyperref}       
\usepackage{url}            
\usepackage{booktabs}       
\usepackage{amsfonts}       
\usepackage{nicefrac}       
\usepackage{microtype}      
\usepackage{lipsum}
\usepackage{graphicx}
\usepackage{multicol,fontenc, calc, indentfirst, fancyhdr, graphicx, lastpage, ifthen, lineno, float, amsmath, setspace, enumitem, mathpazo, booktabs, titlesec, etoolbox, amsthm, hyphenat, natbib, hyperref, footmisc, geometry, caption, url, mdframed, tabto, soul, multirow, microtype, tikz}

\title{DeepInSAR: A Deep Learning Framework for SAR Interferometric Phase Restoration and Coherence Estimation}

\author{
 Xinyao Sun \\
  Multimedia Rsearch Centre\\
  University of Alberta\\
  Edmonton, AB Canada \\
   \And
 Aaron Zimmer \\
 3vGeomatics Inc.\\
 Vancouver, BC Canada \\
  \And
Subhayan Mukherjee \\
Multimedia Rsearch Centre\\
University of Alberta\\
Edmonton, AB Canada \\
\And
Navaneeth Kamballur Kottayil \\
Multimedia Rsearch Centre\\
University of Alberta\\
Edmonton, AB Canada \\
\And
 Parwant Ghuman \\
3vGeomatics Inc.\\
Vancouver, BC Canada \\
\And
Irene Cheng \\
Multimedia Rsearch Centre\\
University of Alberta\\
Edmonton, AB Canada \\
\texttt{locheng@ualberta.ca} \\
\And
}

\begin{document}
\maketitle
\section{Introduction}\label{sec:introduction}
Synthetic Aperture Radar (SAR) is a remote sensing technology, which uses active microwaves to capture ground surface characteristics. An Interferometric SAR (InSAR) image a.k.a interferogram is created from two temporally separated single look complex (SLC) SAR images via the point-wise product of one SLC image with the complex conjugate of the other SLC image. Thus each pixel in an interferogram indicates phase difference between two co-registered SLC images. The phase difference encodes many useful information including deformation of the earth's surface and topographical signals and have been successfully used to obtain the digital elevation model (DEM). InSAR final products are widely used for civil engineering, topography mapping, infrastructure; oil/gas mining; natural hazards monitoring and elevation change detection. In any SAR system, as the satellite circumnavigates earth, SAR sensor launches millions of radar signals toward the earth in the form of microwaves. Then SAR image is represented as a SLC image, which is generated based on radar information echoed back from the ground. However, different ground surface compositions have strong impact on these radar signals. Some are reflected away from the satellite, some are absorbed by non-reflective materials and some are reflected back to the satellite. Signal reflections can be noisy resulting in SAR images with strong speckle noise. Furthermore, temporal and spatial variations between two SLC acquisitions, cause decorrelation, which also affects the interferometric phase \cite{hanssen2001radar}. Noisy SAR images make the interferometric phase filtering step on their output InSAR image more challenging. It is important to point out that the quality of estimated interforgram has direct importance to the whole processing pipeline. The phase noise will affect all subsequent stages from phase-unwrapping operation to the motion signal modelling \cite{zha2008noise}. Therefore, restoration of interferometric phase image becomes a fundamental and crucial step to ensure measurement accuracy in remote sensing. In this regard, the coherence map of interferogram is a crucial indicator showing reliability of the interferometric phase \cite{deledalle2011nl}. Therefore, interferometric phase filtering and coherence estimation are the main focus in this work.

In recent decades,  numerous filtering methods have been proposed. Boxcar is a well-known method because it is straightforward and thus is still widely used nowadays. It simply performs a moving average to estimate the variation of local pixel pattern. In \cite{seymour1994maximum}, authors show that this kind of simple average process is a maximum-likelihood (ML) estimator for interferometric phase and coherence when all involved processes are stationary. Unfortunately, InSAR images are inherently non-stationary because of changing topography and ground displacement. While Boxcar filter can be useful in a flat area, it is not suitable for areas with high slope. In addition, Boxcar outputs are unsatisfactory due to its strong smoothing behaviour caused by simple averaging. In addition to significant phase and coherence estimation error, it is vulnerable to loss of both spatial resolution and fine details. Other classical filters, such as median filter, 2-D Gaussian filter and multi-look processing, also have similar limitations. 

Consequently, researchers started addressing the problem of non-stationary filtering for interferometric phase. Generally speaking, their methods can be categorized into two groups according to whether the filtering is done with or without domain transformation. Lee filter \cite{lee1998new} is a well known classical method working in the original spatial domain. It takes advantage of local fringe morphology modelling with anisotropic filter, which reduces the noise via local statistics and an adaptive window. Researcher in \cite{chao2013refined} introduced an extension of Lee's method by using a minimum mean squared error estimator to exclude singular pixels within a selected direction. Another statistical optimization framework is proposed in \cite{ferraiuolo2004bayesian}, which applies Bayesian estimation in the filtering process. Some adaptive methods are proposed in \cite{vasile2006intensity} \cite{yu2007adaptive}. Vasile et al. designed an intensity-driven adaptive-neighbourhood method for denoising interferometric phase images \cite{vasile2006intensity}. Yu et al. used a low-pass filter along local fringe orientation with an adaptive-contoured-window \cite{yu2007adaptive}. In \cite{wang2016modified} Wang et al. pointed out phase fringe and noise frequency distribution are different, and hence noise can be detected without destroying the fringe signal. There are also some works which estimate maximum posterior probability, as filtered phase image can be obtained by modelling image prior as a Markov random filed (MRF)\cite{baselice2014joint} \cite{ferraiuolo2004bayesian}. However, how to choose appropriate properties as image prior is still an open problem.

Besides studies in the spatial domain, Goldstein filter \cite{goldstein1998radar} is the first frequency domain method with Fourier transformation. One of its extensions \cite{baran2003modification} proposed a technique to preserve the signal in low noise (high coherence) areas by estimating dominant component from local power spectrum of the signal, which also adapts to the local direction of fringes. Other improvements to the Goldstein and Baran filters have been proposed by researchers, who tried to obtain more accurate coherence estimation and overcome the original method's under-filtering issue on low coherent regions \cite{song2014improved} \cite{jiang2014improvement}. A joint method, which uses modified Goldstein and simplified Lee filter, is invented in \cite{wang2011efficient}. This filter particularly focuses on interferometric phase denoising under high dense fringes and low coherent situation. In \cite{cai2008new}, authors pointed out that filtering with adaptive multi-resolution  technique is also necessary because of different sizes and shapes of the interferogram. It improves the filtering quality on fringes via better frequency estimation and invalid estimation correction. 
Inspired by the success of wavelet domain methods on natural image restoration, researchers had started considering wavelet domain for phase noise filtering.  In \cite{lopez2002modeling}, authors first proposed a wavelet domain filter in complex domain (WInPF) based on a complex phase noise model. They proved that phase information and noise can be more easily separated in the wavelet domain. The success of WInPF was of a great importance to lots of subsequent work. \cite{zha2008noise} applied wavelet packets based Wiener filter to further separate phase information in the wavelet packet domain, it achieves superior performance compared to the WInPF filter. In \cite{bian2011interferometric}, Bian and Mercer proposed undecimated wavelet transform by treating image filtering as an estimation problem. Overall, wavelet-domain filters seem to better preserve a good spatial resolution than other methods and have high computational efficiency. Xu et.al \cite{xu2015sparse} introduced a joint denosing filter via simultaneous regularization in the wavelet domain. Phase discontinuities are well preserved through this joint sparse constraint and iterations. 

Following the realization that clean signal phase values are also correlated in temporal domain, in recent years, many methods have started taking the interferogram stack into consideration. Theoretically, it is easier to extract displacement information over a longer period of time. DespecKS \cite{ferretti2011new} introduced a space adaptive processing together with their SqueeSAR procedure that could filter interferometric phase properly by using amplitude SAR images. An adaptive multi-looking filter for airborne single-pass multi-baseline InSAR stacks is proposed in \cite{schmitt2014adaptive}. It achieves faster and more efficient estimation on complex covariance matrices of InSAR data stacks by employing principal component analysis(PCA) based thresholding. A simple and parallelizable filtering approach is proposed in \cite{pepe2015improved} to effectively increase the number of pixels with a high temporal coherence as well as allowing a significant reduction in the overall processing time. 

The idea of Nonlocal filtering is to explore more information from the data itself. In general, images contain repetitive structures such as corners and lines. Those redundant patterns in an image could be analyzed and explored to improve filtering performance. Such nonlocal techniques have been extensively applied for natural image restoration and have gained superior results \cite{buades2005review}. In recent years, more and more studies are deploying nonlocal techniques for SAR data filtering from amplitude images de-specking \cite{deledalle2009iterative,parrilli2012nonlocal,cozzolino2014fast} to interferometric phase denosing \cite{deledalle2011nl,chen2013interferometric,zhu2014improving,sica2018insar,deledalle2015nl}, and InSAR stack multi-temporal processing \cite{su2014two}\cite{sica2015nonlocal}. Compared to the aforementioned methods, although they are promising in some aspects, nonlocal based methods always achieve state-of-the-art results. Nonlocal filtering adapts estimation to the local signal behaviour to deal with non-stationery images like previous approaches, but it not only relies on local neighbourhood of the target pixel, but also takes consideration of the entire image according to the image self-similarity property. The first nonlocal method  applied to interferometric phase filtering was proposed by Deledalle et al. in \cite{deledalle2009iterative}. Both image intensities and interferometric phase information are used to build a nonlocal means model with a probability criterion for estimating pixels.  NL-InSAR \cite{deledalle2011nl}  is the first InSAR application to use a non-local approach for the joint estimation of the reflectivity, interferometric phase and coherence map from a pair of co-registered SLC SAR images. In \cite{chen2013interferometric} and \cite{lin2015nonlocal}, researchers achieve  better results on textural fine details preservation by combining non-local filtering with other conventional natural image processing algorithm, such as pyramidal representation and singular value decomposition. A unified frameworks (NL-SAR) is proposed in \cite{deledalle2015nl} as an extension of NL-InSAR, where an adaptive procedure is carried out to handle very high resolution images. It is able to obtain the best nonlocal estimation with good quality on radar structures and discontinuities reconstruction. Recently,  works on extending and modifying existing image restoration algorithms to suit interferometric phase domain achieve very promising performance. In \cite{wang2016modified}, a modified patch-based locally optimal Wiener (PLOW) method is proposed for interferometric phase filtering that achieves on-par and better results than non-local means. Another famous algorithm, nonlocal block-matching 3D (BM3D) which is widely used for additive white Gaussian noise removing for natural images, also inspired researchers to propose InSAR-BM3D \cite{sica2018insar} which delivered state-of-the-art results for InSAR phase filtering. The method is not able to concurrently estimate phase coherence. Instead, InSAR-BM3D requires coherence map as input and as a result, the performance is likely affected by the accuracy of the coherence estimator.

Deep learning based methods, especially Deep convolutional neural network (CNN) techniques have shown their dominant performance in the past few years on different visual related tasks including image restoration. Milestone works using CNN have shown their ability to outperform almost all conventional algorithms. Built upon the experience of using natural image processing technique to interferometric phase filtering domain, in this work, we propose to integrate a new deep learning based model for SAR interferometric phase restoration and coherence estimation called DeepInSAR. The model is empowered by a set of state-of-the-art deep learning techniques, relying on suitable phase-oriented solutions. We aim to design a more effective joint phase filter and coherence estimator, by learning from the pre-generated training data. We pre-processed InSAR data into a single tensor to do a multi-modal fusion analysis of both phase and amplitude information. A densely connected feature extractor is used to achieve multi-scale feature extraction and fusion. Two subsequent fully connected CNN perform phase filtering and coherence estimation from extracted features respectively. InSAR phase noise can be approximated as a Gaussian. So, we adopt the residual learning strategy, which has been proven by other researchers as effective for removing such type of noise \cite{zhang2017beyond}. Meanwhile, pre-activation and bottleneck \cite{he2016identity}, as well as batch normalization techniques \cite{Ioffe2015BatchNA}, are used to enhance training efficiency and boost the model's performance.

The remainder of the paper is organized as follows, we first define briefly our interferometric phase model in Section 2. Section 3 describes
our DeepInSAR model in detail. Experimental setup, as well as quantitative and qualitative comparison of the performance with three other established methods for both simulated and real data, are presented in Section 4. Future work and conclusion are discussed in Section 5.

\section{Phase Noise Model}
Similar to the classical additive white Gaussian noise (AWGN) degradation mode in natural image restoration problem, an interferometric phase can also be characterized by:
\begin{equation}
\label{eq:phasemodel}
\theta{y} = \theta{x} + v
\end{equation}
which has been validated in \cite{lee1998new}. \(\theta{y}\) denotes the noisy observation, \(\theta{x}\) is clean phase component and \(v\) is the noise with zero mean and \(\sigma\)  standard deviation representing different noise levels. It follows a similar definition in the natural image analysis that clean signals are independent from noise signals. Unfortunately, it is not feasible to directly use natural image processing algorithms in interferometric phase domain, because of branch cuts. According to the SAR interferometric phase calculation, the range of interferometric phase is within \([-\pi, \pi)\), which means that wrapped phase value could jump from negative to positive or positive to negative \(\pi\), and they could represent high-frequency motion signals that should be well preserved. Therefore, in this work, we follow the strategy in \cite{wang2016modified} and \cite{lopez2002modeling} to process the interferometric phase in the complex domain. In other words, the phase noise model could be represented by real and imaginary channels, which are continuous values:

\begin{equation}
\begin{split}
\label{eq:phase_complex_model}
y_{Real} &= cos(\theta_{y}) = Qcos(\theta_{x})+v_{r} = Qx_{Real}+v_{r}\\
y_{Imag} &= sin(\theta_{y}) = Qsin(\theta_{x})+v_{i} = Qx_{Imag}+v_{i}
\end{split}
\end{equation}

The noisy phase observation \(\theta_{y}\) is decomposed into two components \(y_{Real}\) and \(y_{Imag}\). \(v_{r}\) and \(v_{i}\) are AWGN noises in the real and imaginary parts, and they are independent from the underlying clean phase signals \(\theta_{x}\). As analyzed in \cite{wang2016modified} Q is a quality indicator, which is monotonically changing with coherence level. We designed our filtering network based on the above complex phase model. During training, the network learns to filter both real and imaginary parts and then the estimated clean phase $\tilde{\theta}{x}$ could be reconstructed from filtered \(\tilde{x}_{Real}\) and \(\tilde{x}_{Imag}\) as:
\begin{equation}\label{eq:phase_recon}
\tilde{\theta}{x} = arctan\left( \frac{\tilde{x}_{Imag}}{\tilde{x}_{Real}}\right) 
\end{equation}

\section{DeepInSAR}
In this section, we describe our proposed DeepInSAR in detail. The main goal is to establish and validate the idea of using deep learning method to automate and accelerate both interferometric phase filtering and coherence estimation, which are conducted separately in most of existing approaches.
Recently, deep learning studies especially CNNs have been dominating various fields of vision-related tasks. Generally, their excellent performance can be attributed to their powerful feature classification and ability to learn image priors during the training stage. The reason why we choose to use CNN for InSAR filtering and coherence estimation is 1) CNN is effective for capturing spatial feature characterization with a lot of trained parameters, 2) Many achievements in deep learning can be borrowed to benefit better training and generalization, as well as to speed up and improve the output data quality. 3) Powerful GPUs could speed up CNN training and runtime inference. Deep CNN is well suited to be deployed on modern GPUs for parallel computation. All these advantages make deep learning techniques promising for InSAR phase filtering and coherence estimation, where real-time processing and high quality of large resolution radar images are required. 
\begin{figure*}
	\centering
	\includegraphics[width=0.98\linewidth]{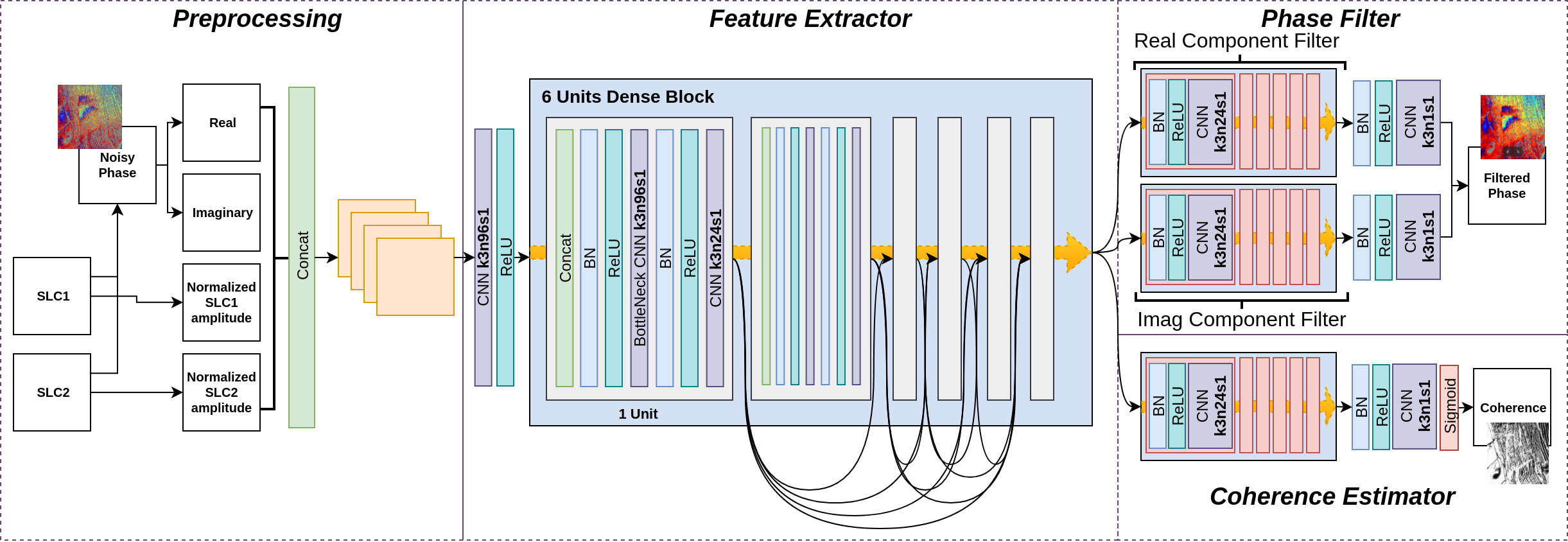}
	\caption{The architecture of the proposed DeepInSAR network}
	\label{fig:network}
\end{figure*}

Fig. \ref{fig:network} illustrates the architecture of the proposed DeepInSAR network. At a high-level, our deep model includes multiple modules for handling different subtasks. The amplitudes and their interferometric phases of two SLC SAR images are combined by concatenating into a single tensor during preprocessing step. The output is subsequently fed into a 
densely connected feature extractor. Dense connectivity helps extract useful features under different scales and composite multi-scale features are suitable for different end tasks \cite{huang2017densely}.   
Two \textit{feature to image} transformations are achieved by sub-networks performing: 1) phase filtering using residual learning strategy \cite{zhang2017beyond} and  2) coherence estimation. 
The model is expected to learn optimal discriminative functions, mapping from noisy observations to both latent clean phase signals and coherence, by a feed-forward neural network.

\subsection{Prepossessing of Radar Data}
Referring to our noise model in Eq.  \ref{eq:phase_complex_model}, we propose to fully utilize all the information from two SLCs rather than only analyzing interferometric phase.  As shown in \textit{Preprocessing Module} in Fig. \ref{fig:network}, the raw input contains two noisy co-registered SLC SAR images \(S_{1}\) and \(S_{2}\). Interferometric phase image \(I\) is calculated as:
\begin{equation}
\label{eq:insar}
I = (A_{S1} \odot\ A_{S2})e^{(\varphi^{S2}-\varphi^{S1})} = A_{I}e^{\Delta\varphi}
\end{equation}
where A is amplitude and \(\varphi\) is phase. In fact, the phases in SLC images look like random noise from one pixel to another because each pixel is a complicated function of scattering features located on the ground surface. However interferometric phase $\Delta\varphi$ represents phase-difference fringes illustrating changes in distance between ground and satellite antenna, which are valuable information needed for InSAR related application but they are often contaminated by noise. Intuitively, we want to incorporate amplitude images, because they usually show recognizable patterns like buildings, mountains, and valleys, which are useful spatial characterizations and hence informative for denoising and coherence estimation. For phase filtering, our DeepInSAR aims to learn a mapping function $\mathcal{F}_{oc}: observation \mapsto clean $. As shown in Eq. \ref{eq:phase_complex_model}, $\mathcal{F}_{oc}$ can include noisy \(y_{Rea1}\), \(y_{Imag}\) and $Q$ as observations. In this work, we further use two SLC's amplitude value to replace the $Q$ in the observations, because we learn from \cite{bamler1998synthetic} that coherence magnitude $|\gamma|$ can be approximated based on two SLC's amplitude:
\begin{equation}
\label{eq:ml-coherence}
|\gamma| = \frac{|\sum^{M}_{m=1}\sum^{N}_{n=1}A_{S1}(m,n)A_{S2}(m,n)|}{\sqrt{\sum^{M}_{m=1}\sum^{N}_{n=1}|A_{S1}(m,n)|^2\sum^{M}_{m=1}\sum^{N}_{n=1}|A_{S2}(m,n)|^2}}
\end{equation} 
where $M,N$ represent estimator window size. This widely used coherence estimator shows a potential mapping $(A_{S1},A_{S2}) \mapsto |\gamma|$. Moreover, As mentioned in Section 2, $Q$ is related to $|\gamma|$. Here we hypothesize that there is a mapping chain $(A_{S1},A_{S2}) \mapsto |\gamma| \mapsto Q$. Hence, instead of using any handcrafted sampling estimator to estimate $Q$. We proposed to use a deep model to approximate the mapping function $\mathcal{F}_{oc}$, in a simplified end-to-end manner by treating both SLC amplitudes together with interferometric phase as input observation to the network. Theoretically, sufficient and well-reasoned input would help the model learn a proper mapping function to estimate latent clean signals more precisely. The same should also supports estimating the quality of signals (coherence). 

Unfortunately, in real world SAR image, the range of amplitude values could be extremely broad, i.e., from $0$ to $10e5$, and the scale of the values also varies across different target sites and types of radar sensor. This is one of the reasons why learning-based studies are not pursued for SAR analysis because using uncontrolled amplitude values to train a deep discriminative model is not effective. In general, the learning-based method requires each input dimension to have a similar distribution with low and controlled variance, which has been suggested by many deep learning studies \cite{glorot2010understanding}\cite{zhang2017beyond}. Unnormalized input data can lead to an awkward loss function topology and place more emphasis on certain parameter gradients resulting in a poor training. Hence, for CNN layer, all the input pixels should be in the same scale. The amplitude values in raw SAR images are not suitable as input data for a deep model. In this work, we introduce an adaptive method to normalize all amplitude values to lie between 0 to 1.
The model saturates potential outliers as well as keeps most dynamic changes in the original image without destroying or cutting off any essential ground characteristics.

\begin{figure*}
	\centering
	
	\includegraphics[width=1\linewidth]{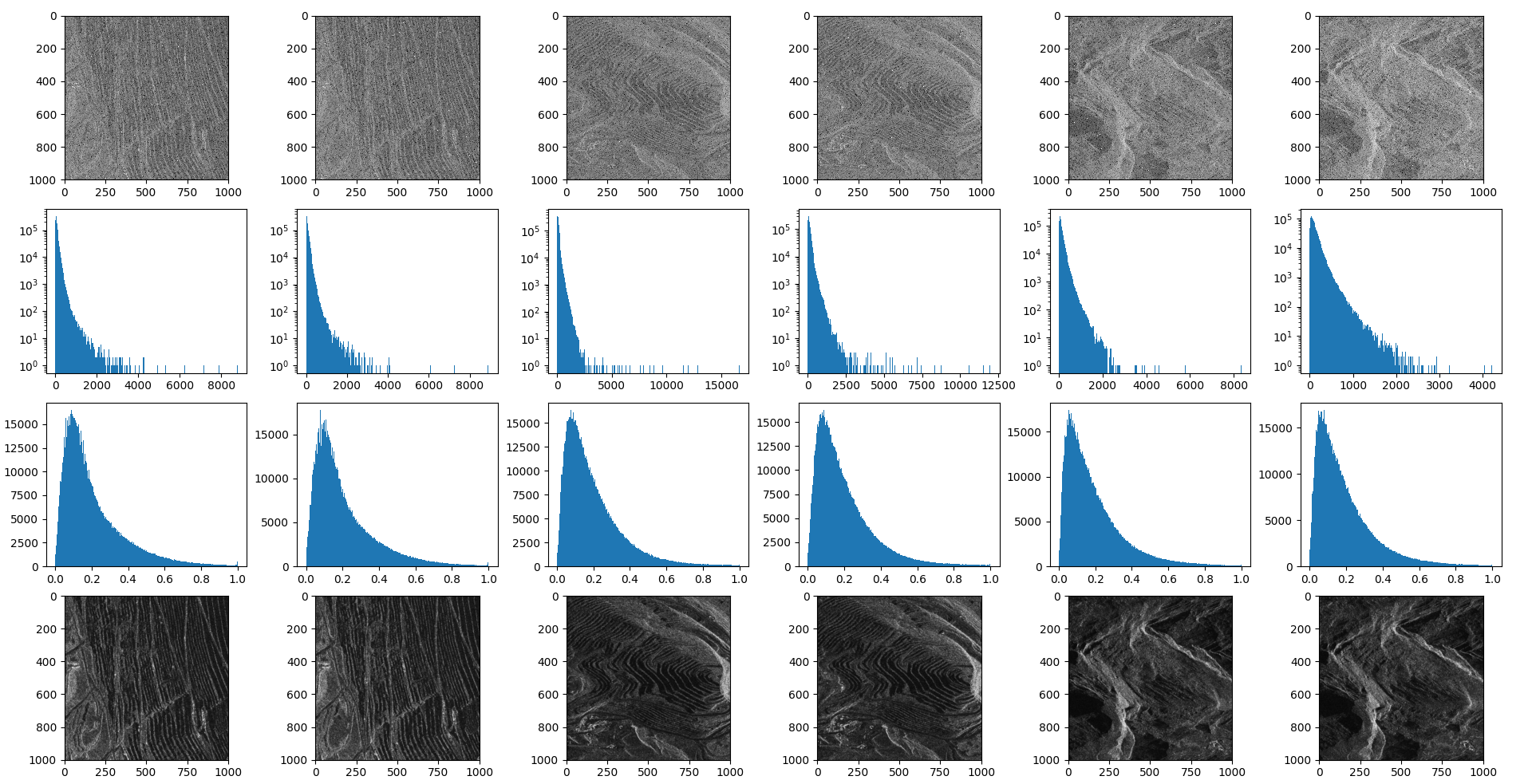}
	
	\caption{Amplitude images selected from three real world site datasets before and after preprocessing. From left to right, it shows Site-A, Site-B, Site-C with two samples for each dataset. (1st row) Raw amplitude images after log transformation for better visualization, (2nd row) their corresponding histograms in log, (3rd row) histograms after proposed normalization and (4th row) corresponding normalized images.}
	\label{fig:normalizaion-preprofessing}
\end{figure*}

Inspired by the classical outliers detector \cite{iglewicz1993detect}, we first calculate median absolute deviation (MAD) for SLC amplitude $A$:
\begin{equation}\label{eq:MAD}
MAD = median(|A_{i}-\tilde{A}|)
\end{equation}
where $\tilde{A}$ is the median of the data. Compared to the average absolute deviation, MAD is less affected by extremes in the distribution tail, and thus it is suitable for real SAR amplitude data.
Next, we transform the data into the modified Z-score  domain:
\begin{equation}\label{eq:mz}
A_{i}^{mz} = \dfrac{0.6745*(A_{i}-\tilde{A})}{MAD} 
\end{equation}
$A^{mz}$ represents each pixel's modified $Z$ score. For outlier detection, researchers commonly use the absolute value of modified $Z$-scores to threshold the data, where data points with $|Z|$ score greater than 3.5 are potential outliers and are ignored \cite{iglewicz1993detect}. By observing amplitude images and their histograms from three real datasets as shown in the 1st and 2nd rows of Fig.\ref{fig:normalizaion-preprofessing}, data points are close to Rayleigh distribution. So simply cutting off according to the modified $Z$ score might cause loss of information located on the right tail of high amplitude values. Although logarithm transformation could help us visualize the images better, there is no fixed base number for all images because they might differ by order of magnitude. In our proposed normalization method, we adopt modified $Z$ score as a transformation function to force all values to be close to 0 first and then all potential outliers will be far from 0 and greater than 3.5. To give a standard input data distribution for training the neural network, we apply hyperbolic tangent $tanh$ non-linear function as:  
\begin{equation}\label{eq:normalizeA}
\hat{A} = \dfrac{1}{2}(tanh(\frac{A^{mz}}{7})+1)
\end{equation}  
to bind all input amplitudes with a controlled variance. A good property of hyperbolic tangent $tanh(x)$ function is that the input value between -1 to 1 will be enhanced and others will be saturated. In our case, we divide $A^{mz}$ by 7 (two times of 3.5) to make the majority of data points lie between -1 to 1. Then ground characteristics could be potentially enhanced after $tanh$ operations. Secondly, data points with relatively high amplitude are still kept on the right tail, and for those extremely high values, likely outliers, are saturated close to 1. Note that, we further normalize the transformed data to the range 0 to 1, because we use a Rectified Linear Unit (ReLU) activation for introducing nonlinearity in the CNN to learn complex features. Non-negative input is recommended to avoid saturated neuron at early training stage when using ReLU activation in the early layers \cite{glorot2011deep}. As shown in the 3rd row in Fig. \ref{fig:normalizaion-preprofessing}, after our proposed data normalization, all amplitude values lie in the range 0 to 1 are properly delivered without losing and breaking essential details. One  can also observe this in the 4th row of Fig.\ref{fig:normalizaion-preprofessing}. The final observation $\textbf{o}$ is a tensor $[y_{real},y_{imag},\hat{A}_{S1},\hat{A}_{S2}]$, and is  the input to DeepInSAR.

\subsection{Filtering with Residual Learning}
Residual learning is designed for solving performance degradation problem on very deep neural networks \cite{szegedy2017inception}. In our interferometric phase filtering, we apply a similar idea but without using too many skip-connections within the network. We only create identity shortcuts for predicting the residuals of both real and imaginary channels. Instead of directly outputting the estimated clean components, the proposed model is trained to predict residuals. The model implicitly filters the latent clean signals with hidden operations within the deep neural network. For each of the real and imaginary channels we have the loss function below:
\begin{equation}
\begin{split}
\mathcal{L(\mathbf{W_{fe},W_{real}})}&=\frac{1}{2}||\mathcal{R}_{real}(\mathbf{o};\mathbf{W_{fe}},\mathbf{W_{real}})\\ &-(y_{real}-x_{real})||^{2}_{F}\\
\mathcal{L(\mathbf{W_{fe},W_{imag}})}&=\frac{1}{2}||\mathcal{R}_{imag}(\mathbf{o};\mathbf{W_{fe}},\mathbf{W_{imag}})\\&-(y_{imag}-x_{imag})||^{2}_{F}
\end{split}
\end{equation}
where, \(\mathbf{W_{fe}}\) , \(\mathbf{W_{real}}\) and \(\mathbf{W_{imag}}\) are the trainable parameters in the model corresponding to feature extractor, real and imaginary channels respectively. For both real and imaginary channels filtering, during the training iterations, our model aims to learn a residual mapping  \({\mathcal{R}}(\mathbf{o})\approx y - \frac{y-v}{Q}\) according to our noise model (Eq.\ref{eq:phasemodel}). Then the clean components can simply be reversed by \(x=y-\mathcal{R}(\mathbf{o})\). \((y,x)\) represents noise-free training sample (patch) pairs. Residual mapping is much easier to learn than the original unreferenced mapping. It has been shown to output excellent results in many low-level vision image inverse restoration problem such as image super-resolution \cite{timofte2014a} and image denoising \cite{zhang2017beyond}. 
To the best of our knowledge, we are the first to use residual learning and CNN to do InSAR phase filtering. The model now learns a residual mapping $\mathcal{R}: observations \mapsto residuals$ on real and imaginary channels respectively. Furthermore, it is known that phase noise variance $\sigma^2_{\theta}$ could be approximated by coherence magnitude $|\gamma|$ \cite{bamler1998synthetic}:
\begin{equation}\label{eq:phase-coherence}
\sigma^2_{\theta} = \dfrac{\pi^2}{3}-\pi arcsin(|\gamma|) + arcsin^2(|\gamma|)-\dfrac{Li_{2}(|\gamma|^2)}{2}
\end{equation}
where $Li_{2}$ is Euler’s dilogarithm. Our input tensor for phase filtering includes two SLC's amplitude, which correlated to coherence magnitude. Hence, our designed observation input is well-reasoned for predicting phase residuals. 


\subsection{Coherence Estimation}
Coherence map is estimated from two co-registered SAR images and is usually used as an indicator of phase quality. Demarcation of image regions based on the degree of contamination (“coherence”) is an important component of the InSAR processing pipeline. 0 coherence denotes complete decorrelation. On the other hand, successful and accurate deformation is measurable with high coherence. Lower quality of interferometry corresponds to decreasing coherence level and increasing level of noise on the phase. Interferometric fringes can only be observed where image coherence prevails. Filtered output is usually combined with coherence map for further processing, because coherence map could tell how much useful signals are potentially within this area. Some of the filtering studies also require coherence map in the filtering process. However, most of them use Maximum Likelihood (ML) estimator (Eq. \ref{eq:ml-coherence}) or its extensions, which are usually significantly biased when using small window sizes. These methods can lose resolution and increase computational cost with large window sizes. Generally speaking, an area on the ground is treated as coherent, when it appears to have similar surface characterization within all images under analysis. However, between two SAR acquisitions, subareas will decorrelate if the land surface is disturbed.  Therefore, CNN is a very good candidate to handle this spatial and non-local based analysis, especially on our input $o$, where almost all necessary information is available for learning the features and capturing mapping functions. During training, the model can learn to capture prior knowledge on all training samples and represent the knowledge as network weights. Intuitively, our method takes a more reliable and robust non-local analysis compared to conventional non-stacked based work, which only considers one interferogram. It is also more time efficient than stacked based method because there is no requirement for doing heavy runtime analysis after training is done. In our model, we have a separate module in DeepInSAR for coherence estimation by using the same features extracted from observations $\textbf{o}$ as shown in Fig. \ref{fig:highlevel}. Because coherence lies in the range [0,1], we calculate sigmoid cross entropy loss, given logits obtained from last convolution layer's output $\mathbf{c} = \mathcal{F}_{oh}(\mathbf{o};\mathbf{W_{fe}},\mathbf{W_{coh}})$:
\begin{equation}
\label{eq:sigmoid_cross}
\begin{split}
\mathcal{L(\mathbf{W_{fe},W_{coh}})} = \textbf{z} * -log(\sigma(\textbf{c})) + (1-\textbf{z})*-log(1-\sigma(\textbf{c})) \\
\text{where } \sigma(\mathbf{c}) = \dfrac{e^\mathbf{c}}{e^\mathbf{c} + 1}
\end{split}
\end{equation}
$\mathbf{z}$ is the reference coherence map that can be pre-calculated by any existing coherence estimator in order to generate training dataset for real images. 

\subsection{Shared Feature Extractor with Dense Connectivity }
\begin{figure}
	\centering
	\includegraphics[width=0.4\linewidth]{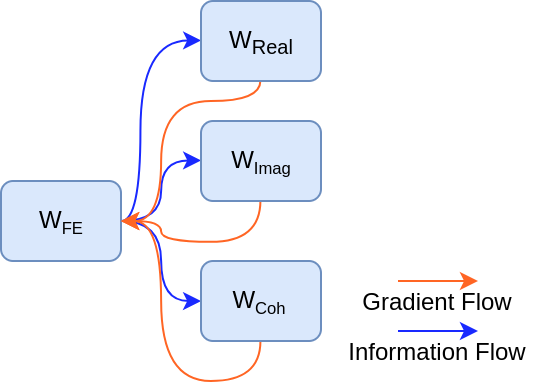}
	\caption{Information and Gradient flow between modules}
	\label{fig:highlevel}
\end{figure}

Natural images exhibit repetitive patterns, such as geometric and photometric similarities, which provide cues to improve the filtering performance. This concept is also valid for InSAR interferometric phase and SAR amplitude images. However, it should be noted that though, CNNs perform well for visual related tasks, it is known that as CNNs become increasingly deep, both input and gradient information can vanish and ``wash out." Recent work ResNet\cite{szegedy2017inception}, \cite{srivastava2015highway} have addressed this problem by building shorter connections between layers close to the input and those close to the output. By doing this, CNNs can be substantially deep but still have accurate performance as well as efficient training. We adopt a dense connected CNN introduced in \cite{huang2017densely} as a shared feature extractor before the real-imaginary filter and coherence estimator. In the single-look interferometric phase, the latent noise level is related to the coherence magnitude \cite{bamler1998synthetic}. A shared feature extractor for both phase filter and coherence estimation modules is expected to capture this relationship in latent space because weights in the feature extractor $\mathbf{W_{fe}}$ are updated based on the gradient feedback back-propagated from both phase residual prediction and coherence estimation as shown in Fig.\ref{fig:highlevel}. During training, the model can encode non-local image prior by updating network parameters according to both phase filter and coherence estimator loss. After training, the model can directly produce filtering and coherence output with a learned discriminative network function without any runtime non-local analysis. 

Furthermore, because of the dense connectivity, our feature extractor follows multi-supervision that  learns to extract common feature parameters for all related subsequent tasks \cite{Ioffe2015BatchNA}. In case of dense connectivity, each layer in the feature extractor is connected to every other layer in a feed-forward manner. During gradient back-propagation, each layer's weight is updated based on all subsequent layers' gradients \cite{huang2017densely}. As shown in Fig. \ref{fig:network}, features extracted by each layer in the feature extractor module of DeepInSAR are based on all preceding layers' output. At the same time, its own output is passed to all subsequent layers as input. In our network, all feature maps extracted at different depth levels are passed to both phase filter and coherence estimator as a single concatenated tensor. 
Note that, as per deep CNNs' working mechanism, early layers extract most detailed and low complexity features with a small perceptual field. With increasing depth, later layers in the feature extractor start extracting high level and complex features with a larger perceptual field. Therefore, a densely connected CNN feature extractor allows each sub-module to perform its own task with multi-scale and muti-complexity features. Our DeepInSAR also achieves a deep supervision by allowing each layer in the feature extractor to have direct access to the gradients from both sub-modules. Dense connectivity guarantees the model to get better feature propagation and enables feature reuse and fusion, which is really important for InSAR phase filtering and coherence estimation. In  real world images, ground data sites contain very different scale level characteristics. That is why most existing methods require user-defined window sizes to extract image characteristics. Therefore, all these methods suffer from the inability to choose a generic optimal window size, and fail to automatically generalize to different data sites. In our case, we use a dense CNN based feature extractor to intelligently select the best multi-level features for subsequent modules. The model is capable of generalizing on phase filtering and coherence estimation for different scale features in one image, as well as performing effectively on new site images.       

\subsection{Teacher-Student Framework}
Based on our findings, the main reason why deep learning techniques have not been pursued widely in InSAR filtering and coherence estimation so far is the lack of ground truth image data (reference without noise) for training such models. For training our DeepInSAR model, we need image pairs as described in Section 3. However, there is no ground truth for real-world InSAR images. Therefore we introduce a teacher-student framework to make it feasible to train DeepInSAR for real-world images.  From the literature, stack-based methods, like PtSel \cite{reza2018accelerating}, always give reliable results. PtSel is an industry level algorithm for coherence estimation and interferometric phase filtering, which searches similar pixels across a stack of interferograms in both spatial and temporal domain.  Despite the accuracy of stack-based methods,  it requires historical SLCs and intensive online parallel searching using a high-end GPU farm, which limits its ability to be integrated into a time-critical InSAR processing chain. The stack-based methods have to wait for several months to collect sufficient data before it can start processing a new site. Although existing stack or non-stack based methods are powerful, most of them require human expert to ensure intermediate output quality because they are incapable of automatically detecting and removing all possible real world noise patterns from InSAR data.
We introduce a deep neural network to replace the manual pre-prccessing, i.e., feature extraction; and post-processing, i.e., quality inspection, with a single intelligent trainable model. 
Similar to training an object classification neural network model, a large human labeled dataset is  required in our approach. Human thus acts as a teacher to teach the model how to classify objects by providing human labeled data. For InSAR phase restoration and coherence estimation, we adopt the PtSel method to create filtered phase images for reference, coherence maps with human tunning and full stack processing to make sure the results are sufficiently reliable. The detail of the PtSel algorithm and its GPU implementation can be found at \cite{reza2018accelerating}\cite{7245704}.  In our approach, PtSel with expert supervision becomes the teacher of the DeepInSAR model, which is a student. We are able to demonstrate that, after training, 1) the student DeepInSAR can generate on-par or even better results than its teacher method - PtSel, using the same test data sets, 2) our model only requires feed-forward inference on a single pair of SLCs, while PtSel requires more than thirty SLCs; and 3) our model can output filtering and coherence results after a one  pass  computation, while PtSel requires back and forward tuning processes and needs the phase unwrapping step, which is time consuming..           

\begin{figure}[h]
	\centering
	\includegraphics[width=0.95\linewidth]{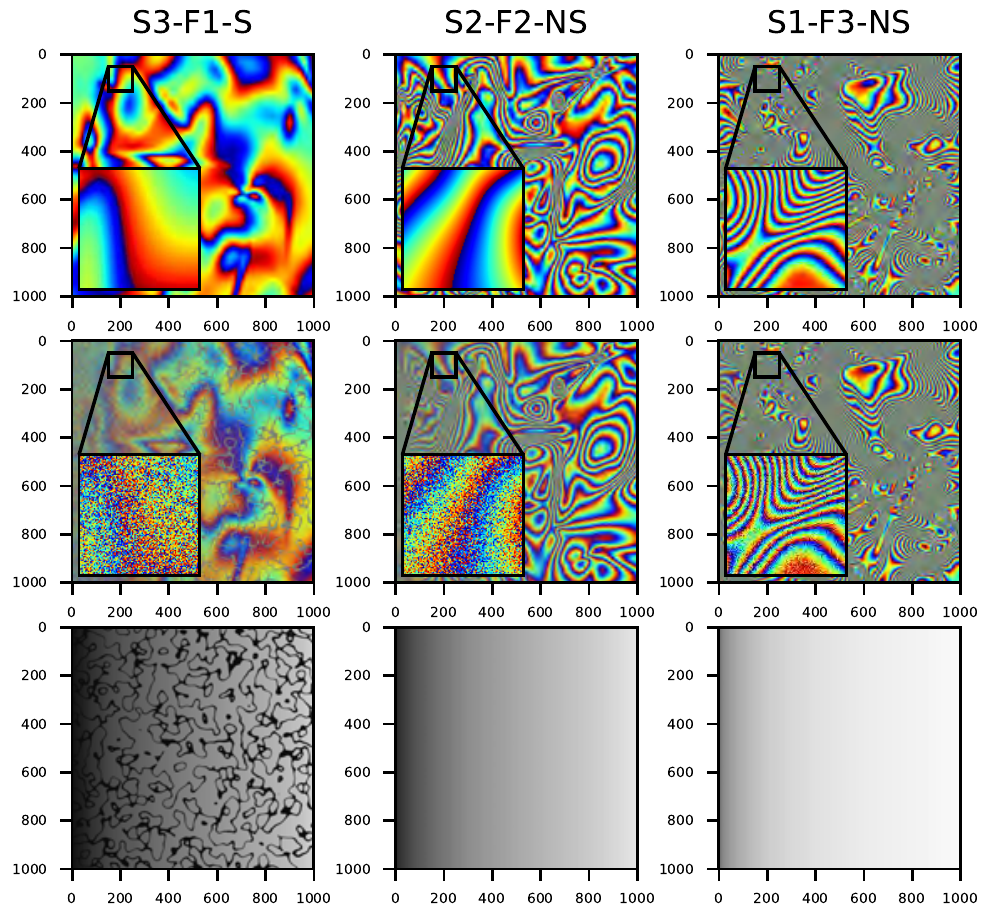}
	\caption{We use S\#-F\#-S or S\#-F\#-NS to name simulation datasets generated using different distortion scenarios: S\# denotes Gaussian level of  base noise S; F\# denotes frequency level of phase fringes F; S and NS mean with or without low amplitude strips respectively. (From let to right) A set of simulated images are selected from S3-F1-S,  S2-F2-NS and S1-F3-NS datasets. First row shows simulated ground truth clean interferometric phase [-$\pi$,$\pi$), second row is the noisy interferometric phase [-$\pi$,$\pi$) - (Blue:-$\pi$; Red:+$\pi$), and third row is coherence (Black:$0$; White:$1$).}
	\label{fig:simsample}
\end{figure}
\section{EXPERIMENTAL ANALYSIS}
We compare our method with a number of other non-stack based methods, which can also perform both phase filtering and coherence estimation. They are 1) boxcar filter 2) NL-SAR \cite{deledalle2015nl} and 3) NL-InSAR \cite{deledalle2011nl}. We used publicly available implementations of these methods found in https://github.com/gbaier/despeckCL. Note that all parameters were set, when applicable, as suggested by the authors of the original papers, or else chosen to optimize the performance. We implemented the proposed DeepInSAR using Tensroflow-GPU 1.10. For a given training dataset, the model was trained on randomly extracted image patches with a size of 128x128. Network parameters were updated using Adam optimizer with a batch size of 64 and 0.001 initial learning rate. The model was trained on two NVIDIA 1080 GPUs for 6 hours with 1.5e6 iterations. To fairly compare the computational time, we executed all methods on the same GPU with an i7-8700K processor and 32GB RAM. It is worth noting that we built and trained our model using common hyper-parameter settings in our experimental setup because the work presented in this paper is mainly for validating the feasibility of using deep learning techniques to do InSAR phase filtering and coherence estimation. It is expected that more extensive hyper-parameter tuning will further improve the performance of our proposed deep model based on the explorations of researchers in \cite{huang2017densely}\cite{timofte2014a}. We conducted experiments using both simulated and real-world data to assess the effectiveness and robustness of the proposed model. In this section, we also discuss learning capacity and generalization ability, which are essential criteria for evaluating a learning model.  

\subsection{Results on Simulation Data}   
In this section, we present quantitative results using simulated data. Simulated data allows us to evaluate the filtered quality in a controlled environment by comparing with the simulated ground truth. Ground truth is treated as an optimal teacher for training our DeepInSAR; we can objectively demonstrate our model's capability to learn proper phase filtering and coherence estimation for new simulated testing images, with ground truth available. The simulation strategy is similar to the work for generating the interferometric phase in \cite{sica2018insar}. Instead of synthesizing a limited known patterns, the additional advantage is to extend the simulation for randomly generated irregular motion signals, ground reflective phenomena, as well as non-stationary noisy conditions. We designed a synthetic InSAR generator to randomly simulate a pair of SLC SAR images with the following procedure:
\vspace{-1em}
\begin{itemize}
	\item Generate first SLC image $S_1$ with 0 phase value. The amplitude value grows from 0.1 to 1 from the left-most column in the image to the right column following a Rayleigh distribution. This leads to a linearly growing of coherence from left to right.
	\item Generate second SLC image $S_2$ by adding random Gaussian bubbles as synthetic motion signals to the phase. The amplitude value is equal to $S_1$'s amplitude value. 
	\item Add random low-value amplitude bands (less than 0.3) on $S1$ and $S2$ to simulate stripe-like low amplitude incoherence areas.
	\item Generate noisy SLCs $S_1^{noisy}$ and $S_2^{noisy}$ by adding independent additive Gaussian noise $v$ to both real and imaginary channels of $S_1$ and $S_2$. 
	\item Calculate clean and noisy interferometric phase $I$ and $I^{noisy}$. 
	\item Calculate ground truth coherence using clean amplitude, phase, and the standard deviation of base noise $v$.
\end{itemize}

Our simulated image generator includes a set of parameters for controlling the complexity of the interferometric phase at different distortion levels. We generated 18 different configurations, by combining (1) three base AWGN levels of $v$ (S1, S2, S3), (2) three fringe frequency levels of phase fringes (F1, F2, F3), and (3) with or without low amplitude strips (S, NS). For example, the dataset, which has a relatively high level of base noise, and low fringe frequency with low amplitude stripes, is denoted by S3-F1-S. Sample images are shown in the first column of Fig \ref{fig:simsample}. We generated 100 samples with 1000x1000 image resolution under each configuration. Half of them were used for training and the rest were for testing. In this experiment, in order to assess the learning capacity and generalization ability of our proposed DeepInSAR model, a single model was trained on all 18 datasets with the noise-free ground truth images (teacher). Because all amplitude stripes and motion signals are randomly generated, all images between training and testing datasets were distinct. Fig. \ref{fig:simsample} shows randomly selected samples from our simulation dataset. Our data generator are inspired by the noise simulation strategy described in \cite{goodman2007speckle}. Basically, we simulate speckle noise by adding uncorrelated zero-mean Gaussian random varibales to the real and imaginary parts of both sythentic SLCs before multiplying them for interferogram generation. To get the ground truth coherence for the simulated interferogram, we make an empirical mapping to it from the standard deviation of those random variables and the ground truth amplitude. This is because increasing the noise will decrease the coherence, and decreasing the amplitude will also decrease the coherence. In this case, each pixel in the generated interferogram is composed of 4 zero-mean Gaussion random variables with identical standard deviation. The source code of our simulator and full resolution simulated samples used in the experiments are available online at https://github.com/Lucklyric/InSAR-Simulator.

\begin{figure}[ht]
	\centering
	\includegraphics[width=0.85\linewidth]{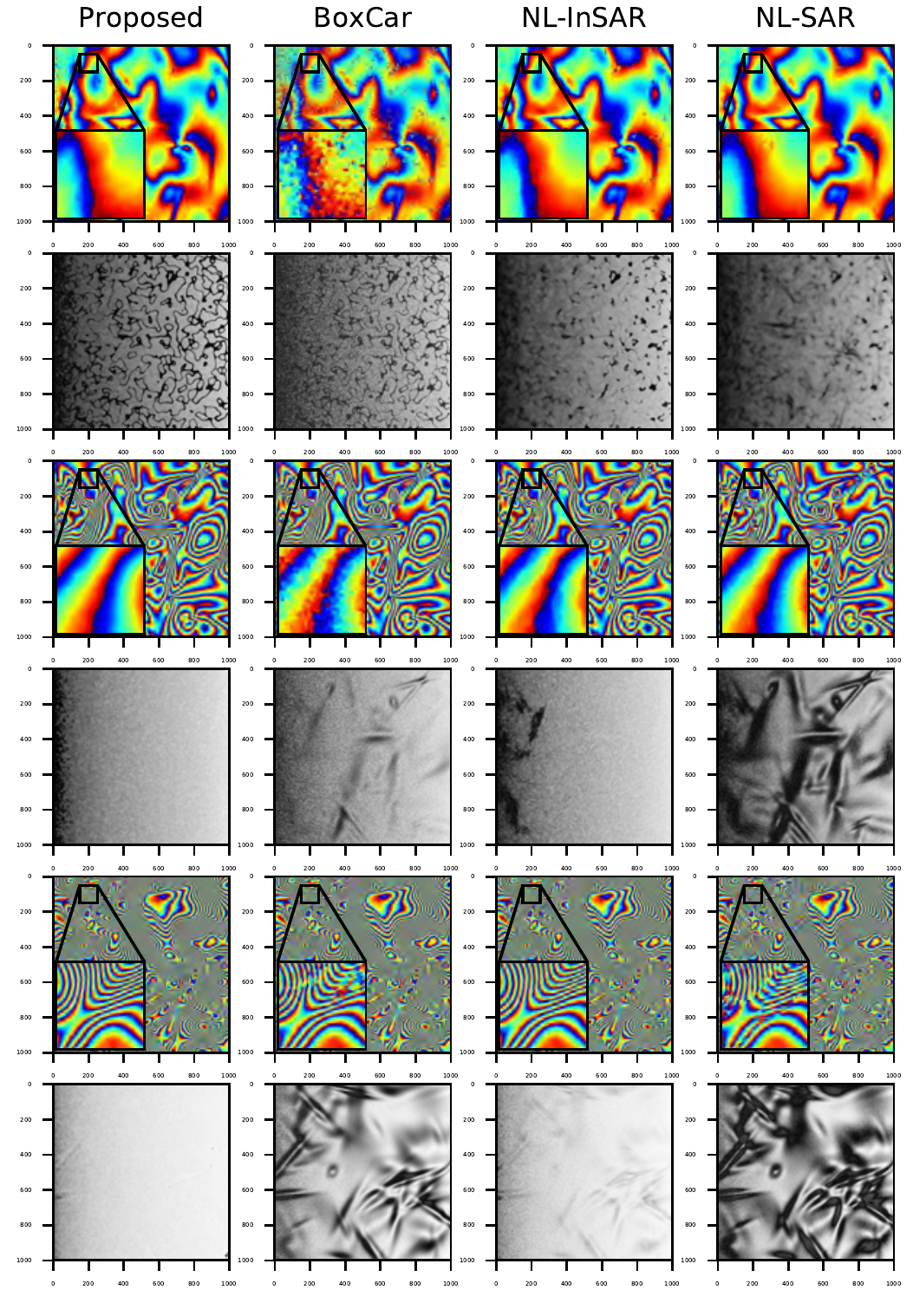}
	\caption{Examples of filtering and coherence estimation results on sample images shown in Fig. \ref{fig:simsample}. Sample images from top to bottom: top two rows are S3-F1-S, middle two rows are S2-F2-NS and last two rows are S1-F3-NS. Visual inspection on filtered outputs from different methods compared to ground truth phase image are given in Fig. \ref{fig:simsample} 1st row. It can be seen that our model can preserve structural details better than others for increasing base noise levels and frequency of fringes (5th row). Our proposed method's coherence estimation is most matched to ground truth (Fig. \ref{fig:simsample} 2nd row), while other methods tend to predict inaccurate results on areas with highly dense fringes or low amplitude stripes.}
	\label{fig:simsampleresults}
\end{figure}
We use objective assessment to evaluate the performance of our method. Our test datasets include 18x50 = 900 simulated images with noisy and ground truth phase images, as well as corresponding coherence indices. The results obtained from Boxcar, NL-InSAR, NL-SAR and our proposed DeepInSAR are compared. We computed both root mean square error (RMSE) in radians, and mean structural similarity map (SSIM) between the filtered phase image and noise-free ground truth to quantitatively evaluate the filtering performance. RMSE is also used to assess coherence estimation.
For visual comparison, sample outputs from Fig. \ref{fig:simsample} are shown in Fig. \ref{fig:simsampleresults}. The quantitative comparison is shown in Table \ref{table:sim_results}. It can be seen that the proposed DeepInSAR significantly outperforms all other methods on most of 18 different distortion levels.  All methods work fairly well on low-level noise and low-level fringe frequency cases. However, with increasing distortion level, all reference methods perform rather poorly. High-level fringe frequency indicates fast-moving areas on the ground. SSIM score indicates that our method preserves excellent detail even on highly dense fringes (F3).  In this case, structural information is one of the most important information that any phase filtering method should preserve, because the performance of subsequent InSAR processing, e.g., phase unwrapping, is heavily affected by the distorted fringe structure. The filtering method should preserve the structural details as much as possible, and our proposed method demonstrates this capability. In particular, when distortion is with high base noise and high fringe frequency, our model only loses insignificant detail especially on relatively low coherent left regions. Although NL-InSAR can guarantee strong noise suppression with detail preservation on high frequency fringes, e.g., (Fig. \ref{fig:simsampleresults} 5th row ), it over-smooths the image when phase distortion level keeps increasing (note the left patch in 5th row of Fig. \ref{fig:simsample} and first row of Fig. \ref{fig:simsampleresults}); fringe structures are washed out when both distortion level and fringe frequency are high (3rd row Fig. \ref{fig:simsampleresults}). Also, NL-SAR can successfully remove noises but its performance is highly dependent on the searching window size. We noticed that when we set a fixed small searching window size, e.g., 25x25, NL-SAR performed well on low frequency fringes at different noise levels (See Table \ref{table:sim_results} two F$1$ columns and 1st row in Fig. \ref{fig:simsampleresults}). However, a larger window size affected its performance on reconstructing high-activity areas. When we adjusted the searching window size back to a smaller size, NL-SAR was able to filter well on those highly dense fringes, but facing under-filtering problem on slow motion areas. During the experiments, we had to manually tune the set of parameters of reference methods in order to get reasonable results.  Their coherence estimators also have similar limitations. BoxCar and NL-SAR tend to output low coherence on fast moving areas (3rd and 6th rows in Fig. \ref{fig:simsampleresults}) and fail to compute correct coherence around low amplitude strips (2nd row in Fig. \ref{fig:simsampleresults}). In comparison, our proposed model's coherence output is most matched to ground truth on all different distortion cases. For instance, all three referenced methods tend to give better results when using (1) a small window size on highly dense fringe areas and  (2) a large window size on low frequency motion. There is no fixed size, which works for all 18 simulated distortion levels.  However, we show that our learning based DeepInSAR works well for all 18 simulated datasets with a single trained model. It has successfully learned the mapping from noisy observations (18 different distortions) to latent clean signals and coherence magnitudes, when we give it proper training samples to explore. Using densely connected feature extractor gives DeepInSAR the ability to intelligently handle multi-scale signal characteristics with a single model. Since the simulated signal patterns are random, therefore simulated motion patterns, noise conditions and low reflective strips, are irregular among all training and testing images. The evaluation output on the testing dataset shows that our trained model does not suffer from the over-fitting issue and only shows a small generalization error. It learns well from the teacher and the model can be generalized to new InSAR data. 
\begin{table}[h]
	\centering
	\caption{Coherence RMSE on 18 different types of Simulation dataset. $S$ denotes Gaussian level of base noise and $F$ represents phase fringes frequency. S and NS mean with and without low amplitude strips respectively.}
	\begin{tabular}{ccccccc}
		\multicolumn{7}{c}{Coherence RMSE}                                                                                                                                                                                                                       \\ \hline
		\multicolumn{3}{c}{\multirow{2}{*}{Sim Configuration}}                                                         & \multicolumn{4}{c}{Methods}                                                                                                             \\ \cline{4-7} 
		\multicolumn{3}{c}{}                                                                                           & Boxcar                      & NL-SAR                      & NL-InSAR                             & Proposed                             \\ \hline
		\multicolumn{1}{|c|}{\multirow{6}{*}{S1}} & \multicolumn{1}{c|}{\multirow{3}{*}{S}}  & \multicolumn{1}{c|}{F1} & \multicolumn{1}{c|}{0.4360} & \multicolumn{1}{c|}{0.4532} & \multicolumn{1}{c|}{0.3827}          & \multicolumn{1}{c|}{\textbf{0.2125}} \\ \cline{3-7} 
		\multicolumn{1}{|c|}{}                    & \multicolumn{1}{c|}{}                    & \multicolumn{1}{c|}{F2} & \multicolumn{1}{c|}{0.5418} & \multicolumn{1}{c|}{0.6356} & \multicolumn{1}{c|}{0.3526}          & \multicolumn{1}{c|}{\textbf{0.1838}} \\ \cline{3-7} 
		\multicolumn{1}{|c|}{}                    & \multicolumn{1}{c|}{}                    & \multicolumn{1}{c|}{F3} & \multicolumn{1}{c|}{0.5321} & \multicolumn{1}{c|}{0.6251} & \multicolumn{1}{c|}{0.3639}          & \multicolumn{1}{c|}{\textbf{0.1850}} \\ \cline{2-7} 
		\multicolumn{1}{|c|}{}                    & \multicolumn{1}{c|}{\multirow{3}{*}{NS}} & \multicolumn{1}{c|}{F1} & \multicolumn{1}{c|}{0.2119} & \multicolumn{1}{c|}{0.3472} & \multicolumn{1}{c|}{\textbf{0.1436}} & \multicolumn{1}{c|}{0.2045}          \\ \cline{3-7} 
		\multicolumn{1}{|c|}{}                    & \multicolumn{1}{c|}{}                    & \multicolumn{1}{c|}{F2} & \multicolumn{1}{c|}{0.5458} & \multicolumn{1}{c|}{0.6515} & \multicolumn{1}{c|}{0.1907}          & \multicolumn{1}{c|}{\textbf{0.1633}} \\ \cline{3-7} 
		\multicolumn{1}{|c|}{}                    & \multicolumn{1}{c|}{}                    & \multicolumn{1}{c|}{F3} & \multicolumn{1}{c|}{0.5444} & \multicolumn{1}{c|}{0.6494} & \multicolumn{1}{c|}{0.2565}          & \multicolumn{1}{c|}{\textbf{0.1564}} \\ \hline
		\multicolumn{1}{|c|}{\multirow{6}{*}{S2}} & \multicolumn{1}{c|}{\multirow{3}{*}{S}}  & \multicolumn{1}{c|}{F1} & \multicolumn{1}{c|}{0.4284} & \multicolumn{1}{c|}{0.4522} & \multicolumn{1}{c|}{0.4136}          & \multicolumn{1}{c|}{\textbf{0.2688}} \\ \cline{3-7} 
		\multicolumn{1}{|c|}{}                    & \multicolumn{1}{c|}{}                    & \multicolumn{1}{c|}{F2} & \multicolumn{1}{c|}{0.4887} & \multicolumn{1}{c|}{0.5564} & \multicolumn{1}{c|}{0.3802}          & \multicolumn{1}{c|}{\textbf{0.2699}} \\ \cline{3-7} 
		\multicolumn{1}{|c|}{}                    & \multicolumn{1}{c|}{}                    & \multicolumn{1}{c|}{F3} & \multicolumn{1}{c|}{0.4784} & \multicolumn{1}{c|}{0.5463} & \multicolumn{1}{c|}{0.3869}          & \multicolumn{1}{c|}{\textbf{0.2774}} \\ \cline{2-7} 
		\multicolumn{1}{|c|}{}                    & \multicolumn{1}{c|}{\multirow{3}{*}{NS}} & \multicolumn{1}{c|}{F1} & \multicolumn{1}{c|}{0.2052} & \multicolumn{1}{c|}{0.3303} & \multicolumn{1}{c|}{\textbf{0.1878}} & \multicolumn{1}{c|}{0.2011}          \\ \cline{3-7} 
		\multicolumn{1}{|c|}{}                    & \multicolumn{1}{c|}{}                    & \multicolumn{1}{c|}{F2} & \multicolumn{1}{c|}{0.4768} & \multicolumn{1}{c|}{0.5664} & \multicolumn{1}{c|}{0.2749}          & \multicolumn{1}{c|}{\textbf{0.2038}} \\ \cline{3-7} 
		\multicolumn{1}{|c|}{}                    & \multicolumn{1}{c|}{}                    & \multicolumn{1}{c|}{F3} & \multicolumn{1}{c|}{0.4766} & \multicolumn{1}{c|}{0.5600} & \multicolumn{1}{c|}{0.3175}          & \multicolumn{1}{c|}{\textbf{0.2166}} \\ \hline
		\multicolumn{1}{|c|}{\multirow{6}{*}{S3}} & \multicolumn{1}{c|}{\multirow{3}{*}{S}}  & \multicolumn{1}{c|}{F1} & \multicolumn{1}{c|}{0.3780} & \multicolumn{1}{c|}{0.3988} & \multicolumn{1}{c|}{0.3834}          & \multicolumn{1}{c|}{\textbf{0.2549}} \\ \cline{3-7} 
		\multicolumn{1}{|c|}{}                    & \multicolumn{1}{c|}{}                    & \multicolumn{1}{c|}{F2} & \multicolumn{1}{c|}{0.4251} & \multicolumn{1}{c|}{0.4836} & \multicolumn{1}{c|}{0.3726}          & \multicolumn{1}{c|}{\textbf{0.2553}} \\ \cline{3-7} 
		\multicolumn{1}{|c|}{}                    & \multicolumn{1}{c|}{}                    & \multicolumn{1}{c|}{F3} & \multicolumn{1}{c|}{0.4244} & \multicolumn{1}{c|}{0.4678} & \multicolumn{1}{c|}{0.3805}          & \multicolumn{1}{c|}{\textbf{0.2591}} \\ \cline{2-7} 
		\multicolumn{1}{|c|}{}                    & \multicolumn{1}{c|}{\multirow{3}{*}{NS}} & \multicolumn{1}{c|}{F1} & \multicolumn{1}{c|}{0.2052} & \multicolumn{1}{c|}{0.2522} & \multicolumn{1}{c|}{0.2086}          & \multicolumn{1}{c|}{\textbf{0.1920}} \\ \cline{3-7} 
		\multicolumn{1}{|c|}{}                    & \multicolumn{1}{c|}{}                    & \multicolumn{1}{c|}{F2} & \multicolumn{1}{c|}{0.4117} & \multicolumn{1}{c|}{0.4904} & \multicolumn{1}{c|}{0.3116}          & \multicolumn{1}{c|}{\textbf{0.1955}} \\ \cline{3-7} 
		\multicolumn{1}{|c|}{}                    & \multicolumn{1}{c|}{}                    & \multicolumn{1}{c|}{F3} & \multicolumn{1}{c|}{0.4207} & \multicolumn{1}{c|}{0.4817} & \multicolumn{1}{c|}{0.3419}          & \multicolumn{1}{c|}{\textbf{0.1998}} \\ \hline
	\end{tabular}
\end{table}
\begin{table}[h]
	\centering
	\caption{Phase RMSE (radians) on 18 different types of Simulation dataset. $S$ denotes Gaussian level of base noise and $F$ represents phase fringes frequency. S and NS mean with and without low amplitude strips respectively.}
	\begin{tabular}{ccccccc}
		\multicolumn{7}{c}{Phase RMSE (radians)}                                                                                                                                                                                                                       \\ \hline
		\multicolumn{3}{c}{\multirow{2}{*}{Sim Configuration}}                                                         & \multicolumn{4}{c}{Methods}                                                                                                             \\ \cline{4-7} 
		\multicolumn{3}{c}{}                                                                                           & Boxcar                      & NL-SAR                               & NL-InSAR                    & Proposed                             \\ \hline
		\multicolumn{1}{|c|}{\multirow{6}{*}{S1}} & \multicolumn{1}{c|}{\multirow{3}{*}{S}}  & \multicolumn{1}{c|}{F1} & \multicolumn{1}{c|}{0.7469} & \multicolumn{1}{c|}{0.8401}          & \multicolumn{1}{c|}{0.8373} & \multicolumn{1}{c|}{\textbf{0.6939}} \\ \cline{3-7} 
		\multicolumn{1}{|c|}{}                    & \multicolumn{1}{c|}{}                    & \multicolumn{1}{c|}{F2} & \multicolumn{1}{c|}{1.0697} & \multicolumn{1}{c|}{1.2012}          & \multicolumn{1}{c|}{0.9572} & \multicolumn{1}{c|}{\textbf{0.7422}} \\ \cline{3-7} 
		\multicolumn{1}{|c|}{}                    & \multicolumn{1}{c|}{}                    & \multicolumn{1}{c|}{F3} & \multicolumn{1}{c|}{1.0699} & \multicolumn{1}{c|}{1.2054}          & \multicolumn{1}{c|}{1.0354} & \multicolumn{1}{c|}{\textbf{0.7890}} \\ \cline{2-7} 
		\multicolumn{1}{|c|}{}                    & \multicolumn{1}{c|}{\multirow{3}{*}{NS}} & \multicolumn{1}{c|}{F1} & \multicolumn{1}{c|}{0.6675} & \multicolumn{1}{c|}{0.7751}          & \multicolumn{1}{c|}{0.7088} & \multicolumn{1}{c|}{\textbf{0.6570}} \\ \cline{3-7} 
		\multicolumn{1}{|c|}{}                    & \multicolumn{1}{c|}{}                    & \multicolumn{1}{c|}{F2} & \multicolumn{1}{c|}{0.9906} & \multicolumn{1}{c|}{1.1015}          & \multicolumn{1}{c|}{0.8284} & \multicolumn{1}{c|}{\textbf{0.6938}} \\ \cline{3-7} 
		\multicolumn{1}{|c|}{}                    & \multicolumn{1}{c|}{}                    & \multicolumn{1}{c|}{F3} & \multicolumn{1}{c|}{0.9623} & \multicolumn{1}{c|}{1.1348}          & \multicolumn{1}{c|}{0.9138} & \multicolumn{1}{c|}{\textbf{0.7261}} \\ \hline
		\multicolumn{1}{|c|}{\multirow{6}{*}{S2}} & \multicolumn{1}{c|}{\multirow{3}{*}{S}}  & \multicolumn{1}{c|}{F1} & \multicolumn{1}{c|}{0.8409} & \multicolumn{1}{c|}{0.8782}          & \multicolumn{1}{c|}{0.9105} & \multicolumn{1}{c|}{\textbf{0.8091}} \\ \cline{3-7} 
		\multicolumn{1}{|c|}{}                    & \multicolumn{1}{c|}{}                    & \multicolumn{1}{c|}{F2} & \multicolumn{1}{c|}{1.1252} & \multicolumn{1}{c|}{1.2319}          & \multicolumn{1}{c|}{1.0859} & \multicolumn{1}{c|}{\textbf{0.8854}} \\ \cline{3-7} 
		\multicolumn{1}{|c|}{}                    & \multicolumn{1}{c|}{}                    & \multicolumn{1}{c|}{F3} & \multicolumn{1}{c|}{1.2096} & \multicolumn{1}{c|}{1.2801}          & \multicolumn{1}{c|}{1.1890} & \multicolumn{1}{c|}{\textbf{0.9593}} \\ \cline{2-7} 
		\multicolumn{1}{|c|}{}                    & \multicolumn{1}{c|}{\multirow{3}{*}{NS}} & \multicolumn{1}{c|}{F1} & \multicolumn{1}{c|}{0.7863} & \multicolumn{1}{c|}{0.8199}          & \multicolumn{1}{c|}{0.8256} & \multicolumn{1}{c|}{\textbf{0.7715}} \\ \cline{3-7} 
		\multicolumn{1}{|c|}{}                    & \multicolumn{1}{c|}{}                    & \multicolumn{1}{c|}{F2} & \multicolumn{1}{c|}{1.0567} & \multicolumn{1}{c|}{1.1687}          & \multicolumn{1}{c|}{0.9854} & \multicolumn{1}{c|}{\textbf{0.8297}} \\ \cline{3-7} 
		\multicolumn{1}{|c|}{}                    & \multicolumn{1}{c|}{}                    & \multicolumn{1}{c|}{F3} & \multicolumn{1}{c|}{1.1251} & \multicolumn{1}{c|}{1.2186}          & \multicolumn{1}{c|}{1.0855} & \multicolumn{1}{c|}{\textbf{0.8785}} \\ \hline
		\multicolumn{1}{|c|}{\multirow{6}{*}{S3}} & \multicolumn{1}{c|}{\multirow{3}{*}{S}}  & \multicolumn{1}{c|}{F1} & \multicolumn{1}{c|}{0.9542} & \multicolumn{1}{c|}{\textbf{0.9332}} & \multicolumn{1}{c|}{0.9648} & \multicolumn{1}{c|}{0.9370}          \\ \cline{3-7} 
		\multicolumn{1}{|c|}{}                    & \multicolumn{1}{c|}{}                    & \multicolumn{1}{c|}{F2} & \multicolumn{1}{c|}{1.1920} & \multicolumn{1}{c|}{1.2657}          & \multicolumn{1}{c|}{1.1883} & \multicolumn{1}{c|}{\textbf{1.0239}} \\ \cline{3-7} 
		\multicolumn{1}{|c|}{}                    & \multicolumn{1}{c|}{}                    & \multicolumn{1}{c|}{F3} & \multicolumn{1}{c|}{1.3080} & \multicolumn{1}{c|}{1.3430}          & \multicolumn{1}{c|}{1.2940} & \multicolumn{1}{c|}{\textbf{1.1156}} \\ \cline{2-7} 
		\multicolumn{1}{|c|}{}                    & \multicolumn{1}{c|}{\multirow{3}{*}{NS}} & \multicolumn{1}{c|}{F1} & \multicolumn{1}{c|}{0.8886} & \multicolumn{1}{c|}{\textbf{0.8672}} & \multicolumn{1}{c|}{0.8976} & \multicolumn{1}{c|}{0.8709}          \\ \cline{3-7} 
		\multicolumn{1}{|c|}{}                    & \multicolumn{1}{c|}{}                    & \multicolumn{1}{c|}{F2} & \multicolumn{1}{c|}{1.1307} & \multicolumn{1}{c|}{1.2203}          & \multicolumn{1}{c|}{1.1159} & \multicolumn{1}{c|}{\textbf{0.9555}} \\ \cline{3-7} 
		\multicolumn{1}{|c|}{}                    & \multicolumn{1}{c|}{}                    & \multicolumn{1}{c|}{F3} & \multicolumn{1}{c|}{1.2398} & \multicolumn{1}{c|}{1.2927}          & \multicolumn{1}{c|}{1.2120} & \multicolumn{1}{c|}{\textbf{1.0259}} \\ \hline
	\end{tabular}
\end{table}
\begin{table}[h]
	\centering
	\caption{Phase SSIM on 18 different types of Simulation dataset. $S$ denotes Gaussian level of base noise and $F$ represents phase fringes frequency. S and NS mean with and without low amplitude strips respectively.}
	\begin{tabular}{ccccccc}
		\multicolumn{7}{c}{Phase SSIM}                                                                                                                                                                                                                           \\ \hline
		\multicolumn{3}{c}{\multirow{2}{*}{Sim Configuration}}                                                         & \multicolumn{4}{c}{Methods}                                                                                                             \\ \cline{4-7} 
		\multicolumn{3}{c}{}                                                                                           & Boxcar                               & NL-SAR                      & NL-InSAR                    & Proposed                             \\ \hline
		\multicolumn{1}{|c|}{\multirow{6}{*}{S1}} & \multicolumn{1}{c|}{\multirow{3}{*}{S}}  & \multicolumn{1}{c|}{F1} & \multicolumn{1}{c|}{0.9424}          & \multicolumn{1}{c|}{0.8897} & \multicolumn{1}{c|}{0.8566} & \multicolumn{1}{c|}{\textbf{0.9511}} \\ \cline{3-7} 
		\multicolumn{1}{|c|}{}                    & \multicolumn{1}{c|}{}                    & \multicolumn{1}{c|}{F2} & \multicolumn{1}{c|}{0.7372}          & \multicolumn{1}{c|}{0.6266} & \multicolumn{1}{c|}{0.7723} & \multicolumn{1}{c|}{\textbf{0.9333}} \\ \cline{3-7} 
		\multicolumn{1}{|c|}{}                    & \multicolumn{1}{c|}{}                    & \multicolumn{1}{c|}{F3} & \multicolumn{1}{c|}{0.6937}          & \multicolumn{1}{c|}{0.5989} & \multicolumn{1}{c|}{0.6888} & \multicolumn{1}{c|}{\textbf{0.9015}} \\ \cline{2-7} 
		\multicolumn{1}{|c|}{}                    & \multicolumn{1}{c|}{\multirow{3}{*}{NS}} & \multicolumn{1}{c|}{F1} & \multicolumn{1}{c|}{\textbf{0.9665}} & \multicolumn{1}{c|}{0.8923} & \multicolumn{1}{c|}{0.9505} & \multicolumn{1}{c|}{0.9585}          \\ \cline{3-7} 
		\multicolumn{1}{|c|}{}                    & \multicolumn{1}{c|}{}                    & \multicolumn{1}{c|}{F2} & \multicolumn{1}{c|}{0.8075}          & \multicolumn{1}{c|}{0.7413} & \multicolumn{1}{c|}{0.8887} & \multicolumn{1}{c|}{\textbf{0.9493}} \\ \cline{3-7} 
		\multicolumn{1}{|c|}{}                    & \multicolumn{1}{c|}{}                    & \multicolumn{1}{c|}{F3} & \multicolumn{1}{c|}{0.7999}          & \multicolumn{1}{c|}{0.7074} & \multicolumn{1}{c|}{0.8117} & \multicolumn{1}{c|}{\textbf{0.9303}} \\ \hline
		\multicolumn{1}{|c|}{\multirow{6}{*}{S2}} & \multicolumn{1}{c|}{\multirow{3}{*}{S}}  & \multicolumn{1}{c|}{F1} & \multicolumn{1}{c|}{0.8898}          & \multicolumn{1}{c|}{0.8590} & \multicolumn{1}{c|}{0.8358} & \multicolumn{1}{c|}{\textbf{0.9122}} \\ \cline{3-7} 
		\multicolumn{1}{|c|}{}                    & \multicolumn{1}{c|}{}                    & \multicolumn{1}{c|}{F2} & \multicolumn{1}{c|}{0.6624}          & \multicolumn{1}{c|}{0.5681} & \multicolumn{1}{c|}{0.6746} & \multicolumn{1}{c|}{\textbf{0.8647}} \\ \cline{3-7} 
		\multicolumn{1}{|c|}{}                    & \multicolumn{1}{c|}{}                    & \multicolumn{1}{c|}{F3} & \multicolumn{1}{c|}{0.5150}          & \multicolumn{1}{c|}{0.4684} & \multicolumn{1}{c|}{0.5202} & \multicolumn{1}{c|}{\textbf{0.7976}} \\ \cline{2-7} 
		\multicolumn{1}{|c|}{}                    & \multicolumn{1}{c|}{\multirow{3}{*}{NS}} & \multicolumn{1}{c|}{F1} & \multicolumn{1}{c|}{0.9221}          & \multicolumn{1}{c|}{0.8902} & \multicolumn{1}{c|}{0.9023} & \multicolumn{1}{c|}{\textbf{0.9312}} \\ \cline{3-7} 
		\multicolumn{1}{|c|}{}                    & \multicolumn{1}{c|}{}                    & \multicolumn{1}{c|}{F2} & \multicolumn{1}{c|}{0.7357}          & \multicolumn{1}{c|}{0.6577} & \multicolumn{1}{c|}{0.7825} & \multicolumn{1}{c|}{\textbf{0.8966}} \\ \cline{3-7} 
		\multicolumn{1}{|c|}{}                    & \multicolumn{1}{c|}{}                    & \multicolumn{1}{c|}{F3} & \multicolumn{1}{c|}{0.6152}          & \multicolumn{1}{c|}{0.5647} & \multicolumn{1}{c|}{0.6398} & \multicolumn{1}{c|}{\textbf{0.8527}} \\ \hline
		\multicolumn{1}{|c|}{\multirow{6}{*}{S3}} & \multicolumn{1}{c|}{\multirow{3}{*}{S}}  & \multicolumn{1}{c|}{F1} & \multicolumn{1}{c|}{0.8026}          & \multicolumn{1}{c|}{0.8168} & \multicolumn{1}{c|}{0.7939} & \multicolumn{1}{c|}{\textbf{0.8349}} \\ \cline{3-7} 
		\multicolumn{1}{|c|}{}                    & \multicolumn{1}{c|}{}                    & \multicolumn{1}{c|}{F2} & \multicolumn{1}{c|}{0.5717}          & \multicolumn{1}{c|}{0.4989} & \multicolumn{1}{c|}{0.5748} & \multicolumn{1}{c|}{\textbf{0.7670}} \\ \cline{3-7} 
		\multicolumn{1}{|c|}{}                    & \multicolumn{1}{c|}{}                    & \multicolumn{1}{c|}{F3} & \multicolumn{1}{c|}{0.3747}          & \multicolumn{1}{c|}{0.3555} & \multicolumn{1}{c|}{0.3919} & \multicolumn{1}{c|}{\textbf{0.6675}} \\ \cline{2-7} 
		\multicolumn{1}{|c|}{}                    & \multicolumn{1}{c|}{\multirow{3}{*}{NS}} & \multicolumn{1}{c|}{F1} & \multicolumn{1}{c|}{0.8570}          & \multicolumn{1}{c|}{0.8722} & \multicolumn{1}{c|}{0.8508} & \multicolumn{1}{c|}{\textbf{0.8824}} \\ \cline{3-7} 
		\multicolumn{1}{|c|}{}                    & \multicolumn{1}{c|}{}                    & \multicolumn{1}{c|}{F2} & \multicolumn{1}{c|}{0.6463}          & \multicolumn{1}{c|}{0.5736} & \multicolumn{1}{c|}{0.6621} & \multicolumn{1}{c|}{\textbf{0.8211}} \\ \cline{3-7} 
		\multicolumn{1}{|c|}{}                    & \multicolumn{1}{c|}{}                    & \multicolumn{1}{c|}{F3} & \multicolumn{1}{c|}{0.4612}          & \multicolumn{1}{c|}{0.4375} & \multicolumn{1}{c|}{0.4938} & \multicolumn{1}{c|}{\textbf{0.7463}} \\ \hline
	\end{tabular}
\end{table}
\begin{figure}[H]
	\centering
	\includegraphics[width=0.98\linewidth]{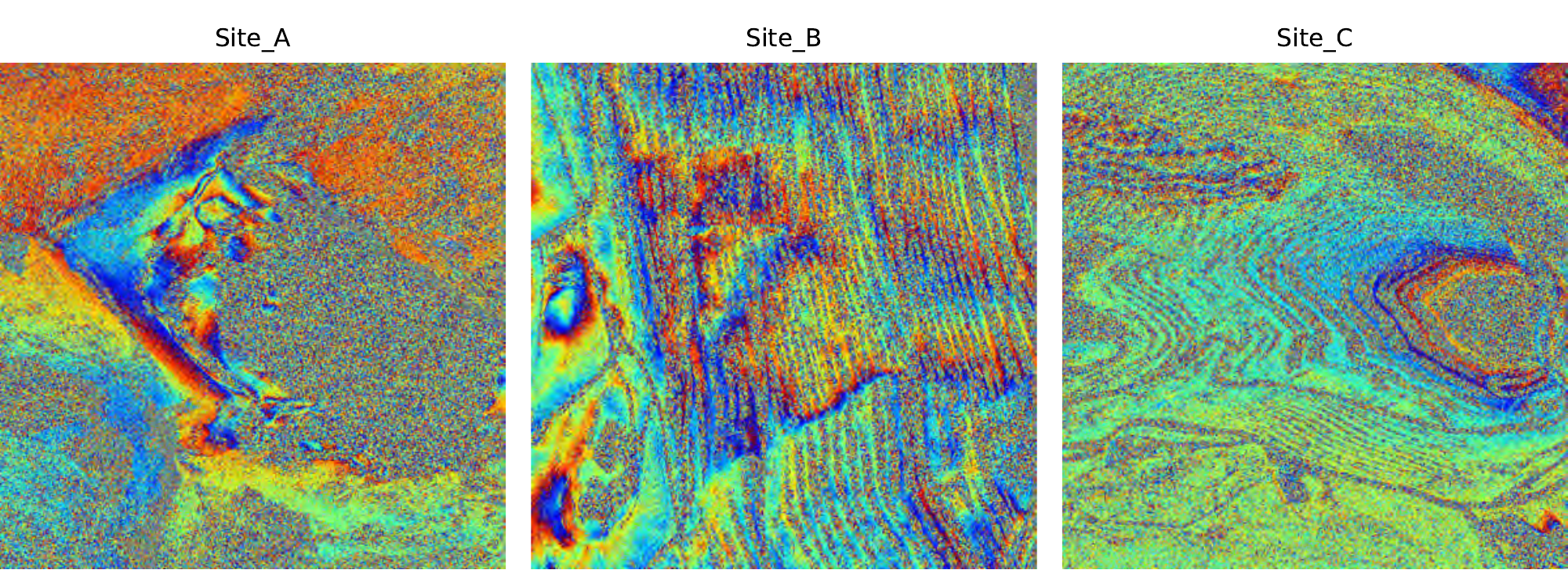}
	\caption{Three representative noisy interferograms (Phase) selected from each of the three real datasets; Blue: -$\pi$; Red:+$\pi$}
	\label{fig:real_noisy}
\end{figure}

\subsection{Results on Real Data}   
Real complex features and noise patterns cannot be fully replicated by simulation data.  However, we can conclude from simulation data experiments that if we can give the model close to clean reference data for teaching DeepInSAR, the model can learn latent mapping from training samples.  As mentioned in Section 3.5, we use PtSel with expert supervision to generate clean reference phases and coherence maps for three real-world datasets captured by TerraSAR-X in StripMap mode \cite{pitz2010terrasar}: 1) Site-A - 27 SLCs  2)  Site-B - 37 SLCs, and 3)  Site-C - 103 SLCs. We used a cropped version of these datasets with size 1000x1000 pixels. 
For coherence estimation, because the window-based PtSel coherence estimator is biased \cite{reza2018accelerating}, we applied binary thresholding $0.5$ on PtSel's coherence output to transform the original regression problem into a classification task. During the inference step, we use coherence estimator's sigmoid output as the confidence level to represent final coherence magnitude. To demonstrate the generalization ability of DeepInSAR on real word InSAR data, we trained the model using images from two sites and tested its robustness on the third site. 

\begin{figure}[h]
	\centering
	\includegraphics[width=0.88\linewidth]{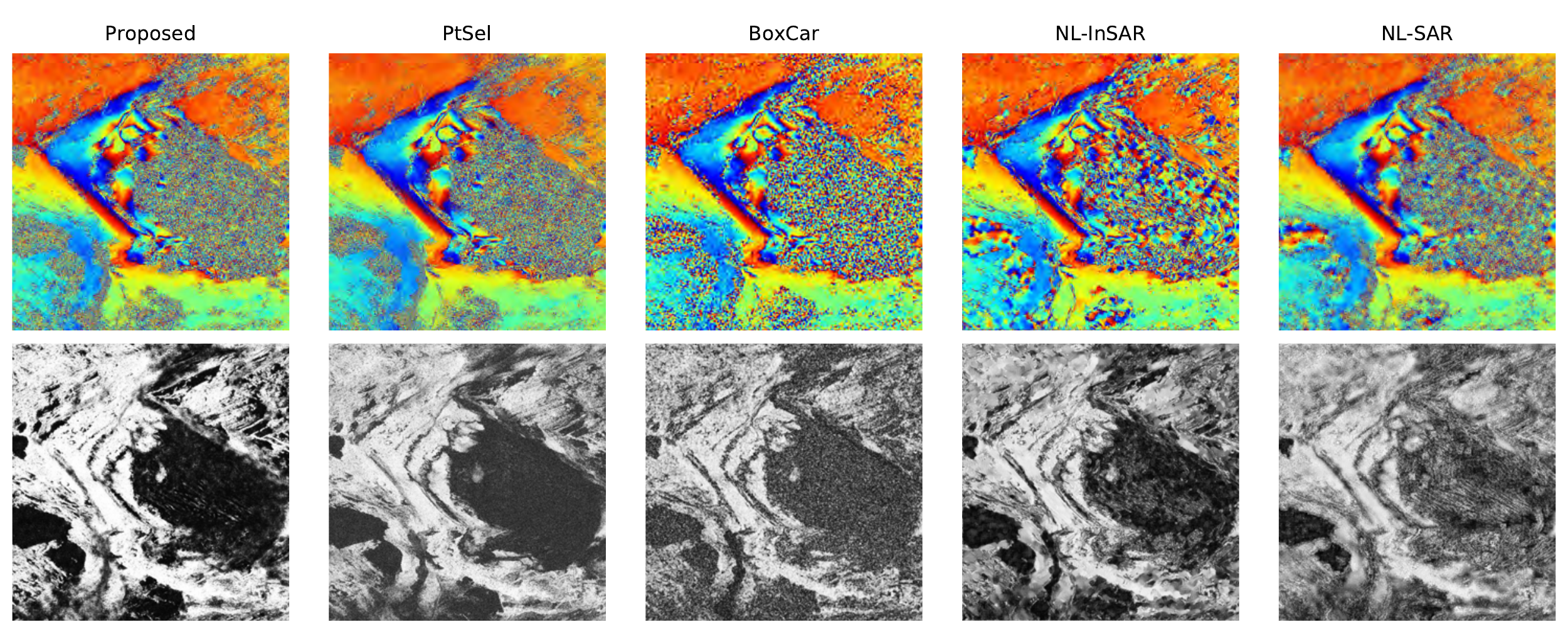}
	\caption{(Top) Filtered images and (Bottom) coherence maps generated by reference methods and trained DeepInSAR model for a  Site-A image.}
	\vspace{2em}
	\centering
	\includegraphics[width=0.88\linewidth]{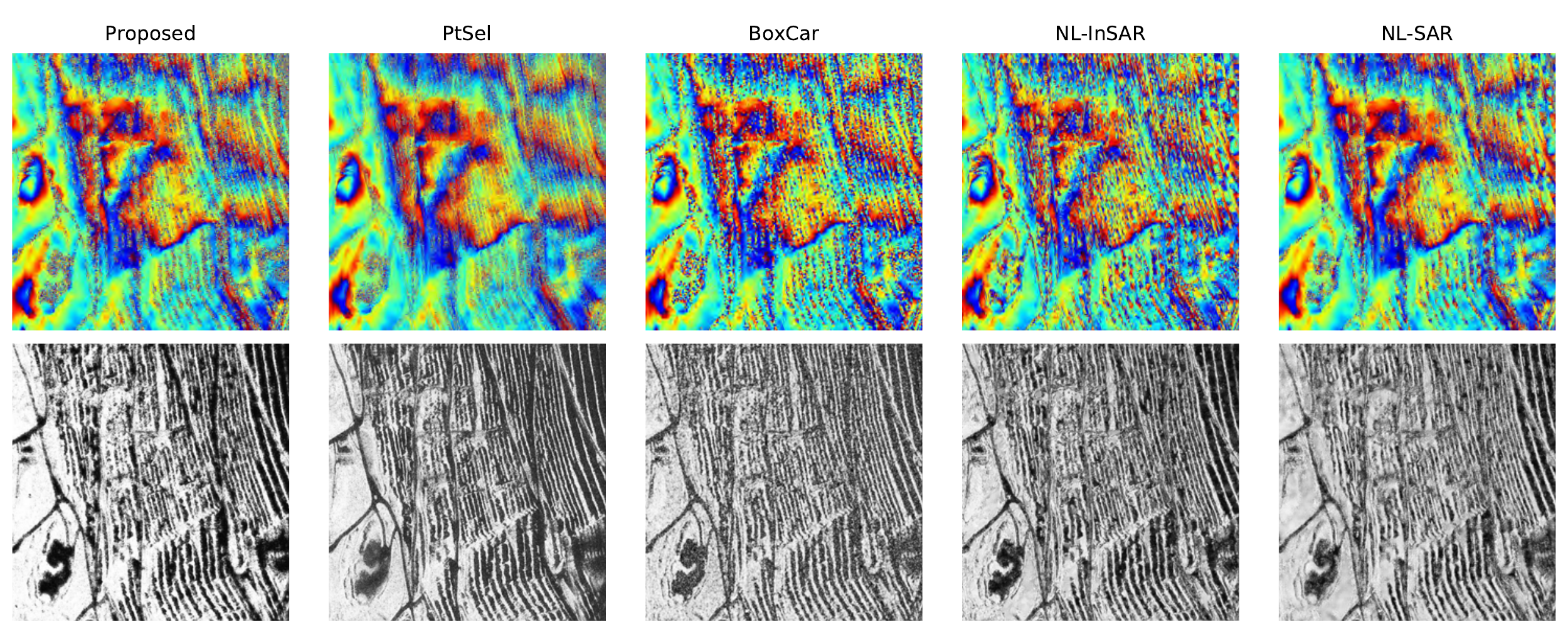}
	\caption{(Top) Filtered images and (Bottom) coherence maps generated by reference methods and trained DeepInSAR model for a  Site-B image.}
	\vspace{2em}
	\centering
	\includegraphics[width=0.88\linewidth]{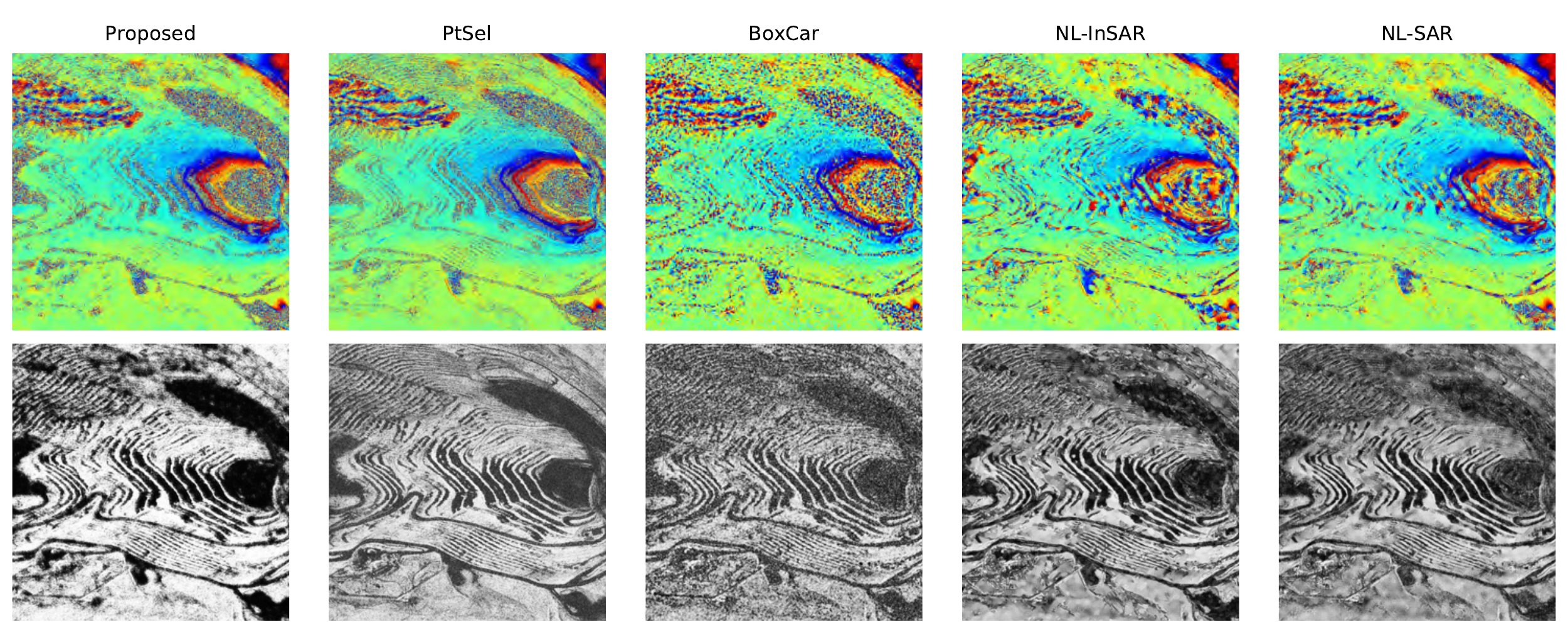}
	\caption{(Top) Filtered images and (Bottom) coherence maps generated by reference methods and trained DeepInSAR model for a  Site-C image.}
	\label{fig:real}
\end{figure}
\begin{table}[]
	\centering
	\caption{Running time T(in seconds) of different methods with image size 1000x1000 }
	\begin{tabular}{|c|c|c|c|c|}
		\hline
		\textbf{}       & \textbf{Boxcar} & \textbf{NL-SAR} & \textbf{NL-InSAR} & \textbf{Proposed} \\ \hline
		\textbf{T(sec)} & 1.16            & 12.77           & 19.36             & 0.46              \\ \hline
	\end{tabular}
	\label{table:running_time}
\end{table}

Three representative interferograms selected from each of the three real datasets are shown in Fig. \ref{fig:real_noisy}. Filtered phases and estimated coherence obtained using BoxCar, NL-InSAR, NL-SAR, PtSel, and our trained DeepInSAR are shown in Fig. \ref{fig:real}. We use qualitative comparison because we do not have noise-free real images for quantity evaluation. 
The boxcar filter tends to blur fringe edges because of its low-pass behaviour and it under-filters near incoherent areas.  Though non-local based NL-SAR and NL-InSAR can provide as sharp and visually appealing filtered phase as DeepInSAR on high coherence areas, on medium and low coherence areas, they tend to flatten the phase and create artifacts in fully noisy areas. Both methods have lower overall variance and less blurring than the boxcar filter, though NL-InSAR has high variance in the estimates between the coherence/amplitude boundaries with streaky artifacts. In NL-SAR's output, there are also artifacts, which are high coherence dots in low coherent area. The limitation is caused by NL-InSAR's numerical instability algorithm and preferential treatment of amplitude, when the amplitude similarities disagree with the phase similarities. Explanation of NL-InSAR's weakness is also discussed in \cite{zimmer2016cuda} and \cite{zhu2018potential}. Comparing to these non-stack based methods, our DeepInSAR offers both strong noise suppression and detail preservation.  Moreover, in Fig. \ref{fig:real}, it is obvious that the output from the three reference models looks very different from the original images of the sites. The comparisons confirm that our trained DeepInSAR model generalizes well to new InSAR data without any human supervision or parameter adjustment, which is required by other methods.  
Furthermore, besides the superior performance compared to other non-stack methods, under a teacher-student framework,  DeepInSAR can achieve results comparable to or better than its teacher method with a learned discriminating neural network. 
The PtSel algorithm (teacher) has several limitations: (1) It relies on temporal information, which means that non-local linear motion can make it hard to pick a neighbourhood suitable for all interferograms, causing under-filtering in these areas. As a result, the algorithm has to wait for many more days of sufficient data before starting the process. (2) It has bias towards filtering results: PtSel looks for similar nearby pixels to perform filtering. If it does not find enough of such pixels, then the filering is towards averaging, giving worse result compared to another pixel which can find lot of similar neighbours.  PtSel's filtering and coherence output is regarded as state-of-the-art in the literature, but it fails to give optimal output across the test input image because of its biased adaptive kernel estimation.  On the other hand, the proposed DeepInSAR successfully distills the knowledge from training samples and generalizes the model to new unseen InSAR images with a simple feed-forward inference, without any human expert supervision, or intensive online searching on a stack of interferograms as required by PtSel. Our DeepInSAR model captures coherence in the fast-moving areas even better than PtSel and produces excellent delineation in the coherence with better contrast, which helps subsequent stages in the InSAR processing pipeline, i.e., when thresholding and weighting are required on the estimated coherence in the phase unwrapping stage. 
With respect to the average running time (T) in seconds, as seen from Table \ref{table:running_time}, the proposed method requires significant less running time than other non-stacked methods because only feed-forward computation is needed after training. After testing different parameter settings (e.g. number of iterations and patch size), reference methods sometimes get better results after running a longer time. However, it is not always the case, which means that these methods have limited potential of full automation without human intervention. The proposed method shows better results with much faster processing time. It is worth mentioning that PtSel outputs used for training and visual comparison are generated using Titan XP GPU farm. This is because PtSel requires high-end GPUs for intensive parallel searching on a stack of SLCs ($>$30). In comparison, our method can run on a consumer level system aforementioned in Section 4, and perform filtering and coherence estimation using only two SLCs. Taking filtering, coherence performance and flexibility into consideration. DeepInSAR is very competitive and suitable for real world InSAR applications.

\section{Conclusion}
In this paper, we propose a learning-based DeepInSAR to address two important research issues: InSAR phase filtering and coherence estimation, in a single process. Our model works well in when using simulated and real data, under different synthetic distortion and real noisy pattern levels. To quantitatively assess DeepInSAR, we designed an InSAR simulator, which can generate motion and noise patterns randomly. We showed that DeepInSAR outperforms existing Non-stack based methods on both tasks. Results show that DeepInSAR can generalize well on new unseen images once it has been trained, and thus can be applied in various real world InSAR applications. We also presented a teacher-student training strategy, which allows DeepInSAR to augment, automate and accelerate existing un-differentiable methods using a differentiable deep neural network. Our trained model can obtain the same or better filtering and coherence estimation results compared to its teacher, requiring less amount of input and achieving higher computational efficiency. Comparing to other non-stack based methds, our model gives better results (1) without any human supervision and (2) with real-time performance. To the best of our knowledge, DeepInSAR is the first work that uses deep neural network to perform InSAR filtering and coherence estimation jointly using both amplitude and phase information of only two co-registered SLC SAR images. In future work, we will investigate how well the DeepInSAR framework can benefit subsequent InSAR analytic stages along the processing pipeline.

\bibliography{DeepInSAR-Arxiv}  

\begin{thebibliography}{10}

\bibitem{hanssen2001radar}
Ramon~F Hanssen.
\newblock {\em Radar interferometry: data interpretation and error analysis},
  volume~2.
\newblock Springer Science \& Business Media, 2001.

\bibitem{zha2008noise}
Xianjie Zha, Rongshan Fu, Zhiyang Dai, and Bin Liu.
\newblock Noise reduction in interferograms using the wavelet packet transform
  and wiener filtering.
\newblock {\em IEEE Geoscience and Remote Sensing Letters}, 5(3):404--408,
  2008.

\bibitem{deledalle2011nl}
Charles-Alban Deledalle, Lo{\"\i}c Denis, and Florence Tupin.
\newblock Nl-insar: Nonlocal interferogram estimation.
\newblock {\em IEEE Transactions on Geoscience and Remote Sensing},
  49(4):1441--1452, 2011.

\bibitem{seymour1994maximum}
MS~Seymour and IG~Cumming.
\newblock Maximum likelihood estimation for sar interferometry.
\newblock In {\em Geoscience and Remote Sensing Symposium, 1994. IGARSS'94.
  Surface and Atmospheric Remote Sensing: Technologies, Data Analysis and
  Interpretation., International}, volume~4, pages 2272--2275. IEEE, 1994.

\bibitem{lee1998new}
Jong-Sen Lee, Konstantinos~P Papathanassiou, Thomas~L Ainsworth, Mitchell~R
  Grunes, and Andreas Reigber.
\newblock A new technique for noise filtering of sar interferometric phase
  images.
\newblock {\em IEEE Transactions on Geoscience and Remote Sensing},
  36(5):1456--1465, 1998.

\bibitem{chao2013refined}
Chin-Fu Chao, Kun-Shan Chen, and Jong-Sen Lee.
\newblock Refined filtering of interferometric phase from insar data.
\newblock {\em IEEE Transactions on Geoscience and Remote Sensing},
  51(12):5315--5323, 2013.

\bibitem{ferraiuolo2004bayesian}
Giancarlo Ferraiuolo and Giovanni Poggi.
\newblock A bayesian filtering technique for sar interferometric phase fields.
\newblock {\em IEEE Transactions on image processing}, 13(10):1368--1378, 2004.

\bibitem{vasile2006intensity}
Gabriel Vasile, Emmanuel Trouv{\'e}, Jong-Sen Lee, and Vasile Buzuloiu.
\newblock Intensity-driven adaptive-neighborhood technique for polarimetric and
  interferometric sar parameters estimation.
\newblock {\em IEEE Transactions on Geoscience and Remote Sensing},
  44(6):1609--1621, 2006.

\bibitem{yu2007adaptive}
Qifeng Yu, Xia Yang, Sihua Fu, Xiaolin Liu, and Xiangyi Sun.
\newblock An adaptive contoured window filter for interferometric synthetic
  aperture radar.
\newblock {\em IEEE Geoscience and Remote Sensing Letters}, 4(1):23--26, 2007.

\bibitem{wang2016modified}
Yang Wang, Haifeng Huang, Zhen Dong, and Manqing Wu.
\newblock Modified patch-based locally optimal wiener method for
  interferometric sar phase filtering.
\newblock {\em ISPRS Journal of Photogrammetry and Remote Sensing}, 114:10--23,
  2016.

\bibitem{baselice2014joint}
Fabio Baselice, Giampaolo Ferraioli, Vito Pascazio, and Gilda Schirinzi.
\newblock Joint insar dem and deformation estimation in a bayesian framework.
\newblock In {\em Geoscience and Remote Sensing Symposium (IGARSS), 2014 IEEE
  International}, pages 398--401. IEEE, 2014.

\bibitem{goldstein1998radar}
Richard~M Goldstein and Charles~L Werner.
\newblock Radar interferogram filtering for geophysical applications.
\newblock {\em Geophysical research letters}, 25(21):4035--4038, 1998.

\bibitem{baran2003modification}
Ireneusz Baran, Michael Stewart, and Peter Lilly.
\newblock A modification to the goldstein radar interferogram filter.
\newblock {\em IEEE Transactions on Geoscience and Remote Sensing},
  41(9):2114--2118, 2003.

\bibitem{song2014improved}
Rui Song, Huadong Guo, Guang Liu, Zbigniew Perski, and Jinghui Fan.
\newblock Improved goldstein sar interferogram filter based on empirical mode
  decomposition.
\newblock {\em IEEE Geoscience and Remote Sensing Letters}, 11(2):399--403,
  2014.

\bibitem{jiang2014improvement}
Mi~Jiang, Xiaoli Ding, Zhiwei Li, Xin Tian, Wu~Zhu, Chisheng Wang, and Bing Xu.
\newblock The improvement for baran phase filter derived from unbiased insar
  coherence.
\newblock {\em IEEE Journal of Selected Topics in Applied Earth Observations
  and Remote Sensing}, 7(7):3002--3010, 2014.

\bibitem{wang2011efficient}
Qingsong Wang, Haifeng Huang, Anxi Yu, and Zhen Dong.
\newblock An efficient and adaptive approach for noise filtering of sar
  interferometric phase images.
\newblock {\em IEEE Geoscience and Remote Sensing Letters}, 8(6):1140--1144,
  2011.

\bibitem{cai2008new}
Bin Cai, Diannong Liang, and Zhen Dong.
\newblock A new adaptive multiresolution noise-filtering approach for sar
  interferometric phase images.
\newblock {\em IEEE Geoscience and Remote Sensing Letters}, 5(2):266--270,
  2008.

\bibitem{lopez2002modeling}
Carlos Lopez-Martinez and Xavier Fabregas.
\newblock Modeling and reduction of sar interferometric phase noise in the
  wavelet domain.
\newblock {\em IEEE Transactions on Geoscience and Remote Sensing},
  40(12):2553--2566, 2002.

\bibitem{bian2011interferometric}
Yong Bian and Bryan Mercer.
\newblock Interferometric sar phase filtering in the wavelet domain using
  simultaneous detection and estimation.
\newblock {\em IEEE Transactions on Geoscience and Remote Sensing},
  49(4):1396--1416, 2011.

\bibitem{xu2015sparse}
Gang Xu, Meng-Dao Xing, Xiang-Gen Xia, Lei Zhang, Yan-Yang Liu, and Zheng Bao.
\newblock Sparse regularization of interferometric phase and amplitude for
  insar image formation based on bayesian representation.
\newblock {\em IEEE Transactions on Geoscience and Remote Sensing},
  53(4):2123--2136, 2015.

\bibitem{ferretti2011new}
Alessandro Ferretti, Alfio Fumagalli, Fabrizio Novali, Claudio Prati, Fabio
  Rocca, and Alessio Rucci.
\newblock A new algorithm for processing interferometric data-stacks: Squeesar.
\newblock {\em IEEE Transactions on Geoscience and Remote Sensing},
  49(9):3460--3470, 2011.

\bibitem{schmitt2014adaptive}
Michael Schmitt and Uwe Stilla.
\newblock Adaptive multilooking of airborne single-pass multi-baseline insar
  stacks.
\newblock {\em IEEE Transactions on Geoscience and Remote Sensing},
  52(1):305--312, 2014.

\bibitem{pepe2015improved}
Antonio Pepe, Yang Yang, Mariarosaria Manzo, and Riccardo Lanari.
\newblock Improved emcf-sbas processing chain based on advanced techniques for
  the noise-filtering and selection of small baseline multi-look dinsar
  interferograms.
\newblock {\em IEEE Transactions on Geoscience and Remote Sensing},
  53(8):4394--4417, 2015.

\bibitem{buades2005review}
Antoni Buades, Bartomeu Coll, and Jean-Michel Morel.
\newblock A review of image denoising algorithms, with a new one.
\newblock {\em Multiscale Modeling \& Simulation}, 4(2):490--530, 2005.

\bibitem{deledalle2009iterative}
Charles-Alban Deledalle, Lo{\"\i}c Denis, and Florence Tupin.
\newblock Iterative weighted maximum likelihood denoising with probabilistic
  patch-based weights.
\newblock {\em IEEE Transactions on Image Processing}, 18(12):2661, 2009.

\bibitem{parrilli2012nonlocal}
Sara Parrilli, Mariana Poderico, Cesario~Vincenzo Angelino, and Luisa
  Verdoliva.
\newblock A nonlocal sar image denoising algorithm based on llmmse wavelet
  shrinkage.
\newblock {\em IEEE Transactions on Geoscience and Remote Sensing},
  50(2):606--616, 2012.

\bibitem{cozzolino2014fast}
Davide Cozzolino, Sara Parrilli, Giuseppe Scarpa, Giovanni Poggi, and Luisa
  Verdoliva.
\newblock Fast adaptive nonlocal sar despeckling.
\newblock {\em IEEE Geoscience and Remote Sensing Letters}, 11(2):524--528,
  2014.

\bibitem{chen2013interferometric}
Runpu Chen, Weidong Yu, Robert Wang, Gang Liu, and Yunfeng Shao.
\newblock Interferometric phase denoising by pyramid nonlocal means filter.
\newblock {\em IEEE Geoscience and Remote Sensing Letters}, 10(4):826--830,
  2013.

\bibitem{zhu2014improving}
Xiao~Xiang Zhu, Richard Bamler, Marie Lachaise, Fathalrahman Adam, Yilei Shi,
  and Michael Eineder.
\newblock Improving tandem-x dems by non-local insar filtering.
\newblock In {\em EUSAR 2014; 10th European Conference on Synthetic Aperture
  Radar; Proceedings of}, pages 1--4. VDE, 2014.

\bibitem{sica2018insar}
Francescopaolo Sica, Davide Cozzolino, Xiao~Xiang Zhu, Luisa Verdoliva, and
  Giovanni Poggi.
\newblock Insar-bm3d: A nonlocal filter for sar interferometric phase
  restoration.
\newblock {\em IEEE Transactions on Geoscience and Remote Sensing},
  56(6):3456--3467, 2018.

\bibitem{deledalle2015nl}
Charles-Alban Deledalle, Lo{\"\i}c Denis, Florence Tupin, Andreas Reigber, and
  Marc J{\"a}ger.
\newblock Nl-sar: A unified nonlocal framework for resolution-preserving
  (pol)(in) sar denoising.
\newblock {\em IEEE Transactions on Geoscience and Remote Sensing},
  53(4):2021--2038, 2015.

\bibitem{su2014two}
Xin Su, Charles-Alban Deledalle, Florence Tupin, and Hong Sun.
\newblock Two-step multitemporal nonlocal means for synthetic aperture radar
  images.
\newblock {\em IEEE Transactions on Geoscience and Remote Sensing},
  52(10):6181--6196, 2014.

\bibitem{sica2015nonlocal}
Francescopaolo Sica, Diego Reale, Giovanni Poggi, Luisa Verdoliva, and
  Gianfranco Fornaro.
\newblock Nonlocal adaptive multilooking in sar multipass differential
  interferometry.
\newblock {\em IEEE Journal of Selected Topics in Applied Earth Observations
  and Remote Sensing}, 8(4):1727--1742, 2015.

\bibitem{lin2015nonlocal}
Xue Lin, Fangfang Li, Dadi Meng, Donghui Hu, and Chibiao Ding.
\newblock Nonlocal sar interferometric phase filtering through higher order
  singular value decomposition.
\newblock {\em IEEE Geoscience and Remote Sensing Letters}, 12(4):806--810,
  2015.

\bibitem{zhang2017beyond}
Kai Zhang, Wangmeng Zuo, Yunjin Chen, Deyu Meng, and Lei Zhang.
\newblock Beyond a gaussian denoiser: Residual learning of deep cnn for image
  denoising.
\newblock {\em IEEE Transactions on Image Processing}, 26(7):3142--3155, 2017.

\bibitem{he2016identity}
Kaiming He, Xiangyu Zhang, Shaoqing Ren, and Jian Sun.
\newblock Identity mappings in deep residual networks.
\newblock In {\em European conference on computer vision}, pages 630--645.
  Springer, 2016.

\bibitem{Ioffe2015BatchNA}
Sergey Ioffe and Christian Szegedy.
\newblock Batch normalization: Accelerating deep network training by reducing
  internal covariate shift.
\newblock In {\em ICML 2015}, 2015.

\bibitem{huang2017densely}
Gao Huang, Zhuang Liu, Laurens Van Der~Maaten, and Kilian~Q Weinberger.
\newblock Densely connected convolutional networks.
\newblock In {\em CVPR}, volume~1, page~3, 2017.

\bibitem{bamler1998synthetic}
Richard Bamler and Philipp Hartl.
\newblock Synthetic aperture radar interferometry.
\newblock {\em Inverse problems}, 14(4):R1, 1998.

\bibitem{glorot2010understanding}
Xavier Glorot and Yoshua Bengio.
\newblock Understanding the difficulty of training deep feedforward neural
  networks.
\newblock In {\em Proceedings of the thirteenth international conference on
  artificial intelligence and statistics}, pages 249--256, 2010.

\bibitem{iglewicz1993detect}
Boris Iglewicz and David~Caster Hoaglin.
\newblock {\em How to detect and handle outliers}, volume~16.
\newblock Asq Press, 1993.

\bibitem{glorot2011deep}
Xavier Glorot, Antoine Bordes, and Yoshua Bengio.
\newblock Deep sparse rectifier neural networks.
\newblock In {\em Proceedings of the fourteenth international conference on
  artificial intelligence and statistics}, pages 315--323, 2011.

\bibitem{szegedy2017inception}
Christian Szegedy, Sergey Ioffe, Vincent Vanhoucke, and Alexander~A Alemi.
\newblock Inception-v4, inception-resnet and the impact of residual connections
  on learning.
\newblock In {\em AAAI}, volume~4, page~12, 2017.

\bibitem{timofte2014a}
Radu Timofte, Vincent De~Smet, and Luc Van~Gool.
\newblock A+: Adjusted anchored neighborhood regression for fast
  super-resolution.
\newblock In {\em Asian Conference on Computer Vision}, pages 111--126.
  Springer, 2014.

\bibitem{srivastava2015highway}
Rupesh~Kumar Srivastava, Klaus Greff, and J{\"u}rgen Schmidhuber.
\newblock Highway networks.
\newblock {\em arXiv preprint arXiv:1505.00387}, 2015.

\bibitem{reza2018accelerating}
Tahsin Reza, Aaron Zimmer, Jos{\'e} Manuel~Delgado Blasco, Parwant Ghuman,
  Tanuj~Kr Aasawat, and Matei Ripeanu.
\newblock Accelerating persistent scatterer pixel selection for insar
  processing.
\newblock {\em IEEE Transactions on Parallel and Distributed Systems},
  29(1):16--30, 2018.

\bibitem{7245704}
T.~Reza, A.~Zimmer, P.~Ghuman, T.~k.~Aasawat, and M.~Ripeanu.
\newblock Accelerating persistent scatterer pixel selection for insar
  processing.
\newblock In {\em 2015 IEEE 26th International Conference on
  Application-specific Systems, Architectures and Processors (ASAP)}, pages
  49--56, July 2015.

\bibitem{goodman2007speckle}
Joseph~W Goodman.
\newblock {\em Speckle phenomena in optics: theory and applications}.
\newblock Roberts and Company Publishers, 2007.

\bibitem{pitz2010terrasar}
Wolfgang Pitz and David Miller.
\newblock The terrasar-x satellite.
\newblock {\em IEEE Transactions on Geoscience and Remote Sensing},
  48(2):615--622, 2010.

\bibitem{zimmer2016cuda}
Aaron Zimmer and Parwant Ghuman.
\newblock Cuda optimization of non-local means extended to wrapped gaussian
  distributions for interferometric phase denoising.
\newblock {\em Procedia Computer Science}, 80:166--177, 2016.

\bibitem{zhu2018potential}
Xiao~Xiang Zhu, Gerald Baier, Marie Lachaise, Yilei Shi, Fathalrahman Adam, and
  Richard Bamler.
\newblock Potential and limits of non-local means insar filtering for tandem-x
  high-resolution dem generation.
\newblock {\em Remote Sensing of Environment}, 218:148--161, 2018.

\end{thebibliography}


\end{document}